\documentclass[letterpaper,twocolumn,10pt]{article}
\usepackage{usenix2019_v3}

\usepackage{tikz}
\usepackage{amsmath}
\usepackage{titling}

\usepackage{algorithm}
\usepackage{amsthm}
\usepackage{amssymb}
\usepackage{multirow}
\usepackage{bigstrut}
\usepackage{comment}
\usepackage[noend]{algpseudocode}

\usepackage{ifthen}
\usepackage{booktabs}
\usepackage{svg}

\usepackage[T1]{fontenc}
\usepackage{inconsolata}

\usepackage{graphicx}
\usepackage{subcaption}
\usepackage{textcomp}
\usepackage{tabularx}
\usepackage{enumitem}
\usepackage{cite}
\usepackage{listings}
\usepackage{filecontents}
\usepackage{xcolor}
\usepackage[normalem]{ulem}
\usepackage{multirow}
\usepackage{xspace}
\usepackage{url}
\usepackage[binary-units]{siunitx}
\usepackage{svg} 
\makeatletter
\def\do@url@hyp{\do\-\do\_}
\makeatother

\definecolor{mygreen}{rgb}{0,0.6,0}
\definecolor{mygray}{rgb}{0.5,0.5,0.5}
\definecolor{mymauve}{rgb}{0.58,0,0.82}
\definecolor{myred}{rgb}{0.79,0.15,0.15}
\usepackage[dvipsnames]{xcolor}

\lstdefinestyle{mypython}{
  language=Python,
  backgroundcolor=\color{white},   
  basicstyle=\scriptsize\ttfamily,        
  breakatwhitespace=false,         
  breaklines=true,                 
  captionpos=b,                    
  commentstyle=\color{mygreen},
  deletekeywords={...},            
  escapeinside={\%*}{*)},          
  extendedchars=true,              
  firstnumber=1,                
  frame=no,                    
  keepspaces=true,                 
  keywordstyle=\color{mymauve},       
  language=Octave,                 
  morekeywords={DIRECT, PERIODIC, IMMEDIATE, BY_TIME, EVERY_OBJ},            
  numbers=left,                    
  numbersep=5pt,                   
  numberstyle=\tiny\color{mygray}, 
  rulecolor=\color{black},         
  showspaces=false,                
  showstringspaces=false,          
  showtabs=false,                  
  stepnumber=1,                    
  stringstyle=\color{myred},     
  tabsize=4,                    
  title=\lstname                   
}

\lstdefinestyle{trigger}{
  language=C++,
  basicstyle=\scriptsize\ttfamily,
  breaklines=true,
  commentstyle=\color{mygreen},
  captionpos=b,
  keywordstyle=\color{blue},
  morekeywords={abstract, string},
  showspaces=false,  
  showstringspaces=false,          
  showtabs=false,  
  tabsize=2,                    
  title=\lstname  
}

\lstdefinestyle{mycpp}{
  language=C++,
  basicstyle=\scriptsize\ttfamily,
  breaklines=true,
  commentstyle=\color{mygreen},
  keywordstyle=\color{blue},
  firstnumber=1,   
  numbers=left,
  numbersep=5pt,
  captionpos=b,
  showspaces=false,  
  numberstyle=\tiny\color{mygray},
  showstringspaces=false,          
  stringstyle=\color{myred},     
  showtabs=false,  
  tabsize=2,                    
  title=\lstname  
}

\lstdefinestyle{interface}{
  language=C++,
  basicstyle=\footnotesize\ttfamily,
  breaklines=true,
  keywordstyle=\color{blue},
  captionpos=b,                    
}

\newfloat{lstfloat}{htbp}{lop}  
\floatname{lstfloat}{Listing}

\newcommand{\subtitle}[1]{\noindent\textbf{#1}} 

\newcommand{\PHM}[1]{\vspace{.2em}
\noindent\textbf{#1}\hspace{.5em}} 

\newcommand{\todo}[1]{\noindent\textcolor{blue}{TODO: #1}}
\newcommand{\fix}[1]{\noindent\textcolor{PineGreen!70!black}{#1}}
\newcommand{\yewang}[1]{\noindent\textcolor{RedOrange}{#1}}
\newcommand{\zzx}[1]{\noindent\textcolor{RoyalBlue}{#1}}

\newcommand{\SysNameTitle}{\texorpdfstring{\textbf{J{\large ANUS}}}{Janus}}
\newcommand{\SysName}{\texorpdfstring{\textsc{Janus}\xspace}{Janus}}

\newcommand{\reqbox}[2]{%
  \par\vspace{0.3em}\noindent
  \fbox{\parbox{\dimexpr\linewidth-2\fboxsep-2\fboxrule\relax}{%
    \emph{\textbf{#1}: #2}%
  }}%
  \par\vspace{0.3em}%
}

\newboolean{exclude_appendix}
\setboolean{exclude_appendix}{false}

\graphicspath{{figures/}}

\begin{document}

\date{}


\title{\Large \bf \texorpdfstring{\SysNameTitle}{Janus}: Disaggregating Attention and Experts for Scalable MoE Inference}

\author{
{\rm Zhexiang Zhang\textsuperscript{1*},
Ye Wang\textsuperscript{1,2*},
Yumiao Zhao\textsuperscript{3},
Jiayu Xiao\textsuperscript{1},
Qianjing Yang\textsuperscript{1},} \\
{\rm Xiangyu Wang\textsuperscript{2},
Jingzhe Jiang\textsuperscript{1},
Qizhen Weng\textsuperscript{2},
Ruichuan Chen\textsuperscript{4},
Shaohuai Shi\textsuperscript{5},} \\
{\rm Yin Chen\textsuperscript{2},
Minchen Yu\textsuperscript{1\textdagger}} \\[1.5ex] 
{\itshape
\textsuperscript{1}The Chinese University of Hong Kong, Shenzhen} \\
{\itshape
\textsuperscript{2}Institute of Artificial Intelligence (TeleAI), China Telecom} \\
{\itshape
\textsuperscript{3}Shenzhen Loop Area Institute \quad
\textsuperscript{4}Nokia Bell Labs \quad
\textsuperscript{5}Harbin Institute of Technology, Shenzhen}
}

\maketitle

\makeatletter
\def\blfootnote{\xdef\@thefnmark{}\@footnotetext}
\makeatother

\blfootnote{* Equal contribution.}
\blfootnote{\textdagger\ Corresponding author: Minchen Yu (yuminchen@cuhk.edu.cn).}

\begin{abstract}

Serving large Mixture-of-Experts (MoE) models is challenging because of their substantial resource demands and highly dynamic inference workloads. Most existing MoE inference systems deploy the entire model as a monolithic unit, forcing attention and MoE layers to share the same resource configuration despite their different requirements. Such coarse-grained provisioning leads to resource inefficiency and suboptimal performance. We present \SysName, a scalable and resource-efficient MoE inference system built around three key principles. First, \SysName disaggregates attention and MoE layers onto separate GPU worker pools, enabling independent resource provisioning for these two layer types, and employs an adaptive two-phase communication mechanism for low-latency data exchange. Second, because MoE layers are largely memory-bound and their latency is dominated by the activated-expert counts, \SysName introduces a microsecond-scale activation scheduler that balances per-layer activated experts across MoE instances to reduce the inference latency. 
Third, \SysName employs a fine-grained, SLO-aware resource scaling scheme that jointly optimizes the attention-side and MoE-side resources, together with the expert placement, to minimize the resource cost under token-level SLO constraints. Evaluation shows that \SysName improves per-GPU throughput by up to 4.7$\times$ over the state-of-the-art systems while satisfying token-level latency SLOs.

\end{abstract}
\section{Introduction}
\label{sec:intro}

Recent advances in Large Language Models (LLMs) have driven widespread adoption across diverse domains. As mainstream LLMs scale to ever-larger parameter sizes~\cite{zoph2022emergent}, the Mixture-of-Experts (MoE) has emerged as a dominant architecture, which effectively expands model capacity without proportionally increasing per-token computation~\cite{qwen3,deepseekv3,moelighting,moe-infinity}. Compared with dense LLMs, MoE models retain the attention layers but replace each Feed-Forward Network (FFN) layer with an MoE layer composed of multiple FFNs as experts. These experts are sparsely activated at inference time---only a small subset is invoked per token---and thus keeping per-token computation bounded.

Serving large MoE models for online workloads introduces a fundamental tradeoff between resource efficiency and latency Service Level Objectives (SLOs). 
Real-world LLM services face highly dynamic demand, including fluctuating request arrival rates and highly variable input and output token lengths~\cite{llmtrace2024, burstGPT_arxiv24,megascale,blitzscale,yu2025lambdascaleenablingfastscaling,serverlessllm}. 
Meeting token-level latency SLOs such as Time-Per-Output-Token (TPOT) requires provisioning sufficient GPUs to absorb workload bursts, but this leaves substantial resources idle during low-demand periods. This inefficiency is more pronounced for MoE models because expert parameters dominate the model memory footprint~\cite{moelighting,deepseekv3}. To maintain low latency, most experts need to remain resident in GPU memory even though only a small subset is activated for each token, driving the resource needs far beyond a small GPU deployment (e.g., hosting DeepSeek-V3 requires at least 16 H100 GPUs~\cite{deepseekv3}). Therefore, achieving scalable MoE inference requires meeting stringent token-level SLOs while minimizing resource over-provisioning.

Despite recent progress, achieving scalable MoE inference remains an open challenge. Most existing systems serve the full MoE model as a monolithic instance and manage resources at the granularity of model instances~\cite{moe-infinity,lina,moelighting}. This coarse-grained design overlooks the heterogeneous resource demands of attention and MoE layers: MoE layers are more memory-intensive, and their latency depends strongly on the number of activated experts per GPU (see \S\ref{sec:characteristics}). It forces attention and MoE layers to use the same resource configuration, such as identical parallelism degrees, leading to either over-provisioning or degraded performance. Recent systems such as MegaScale-Infer~\cite{megascale} disaggregate attention and MoE layers onto separate clusters. However, they still offer limited support for fine-grained resource scaling and expert management, often relying on a static expert-to-GPU placement, e.g., pinning fixed sets of experts to dedicated GPUs. Consequently, they remain resource-inefficient under dynamic workloads.

An ideal MoE inference system should satisfy three key requirements. First, it should support \emph{independent resource provisioning for attention and MoE layers}, allowing each to adopt a configuration tailored to its own demand. 
Second, it should \emph{balance the number of activated experts across GPUs to maximize the efficiency of MoE execution}, which is largely memory-bound and dominates the end-to-end resource footprint.
Third, it should enable \emph{fine-grained elasticity under SLOs}, incrementally adjusting capacity and expert placement to ensure that the resulting configuration meets token-level latency constraints with the minimal resource cost.

Guided by these requirements, we present \SysName, a scalable and resource-efficient MoE inference system.
\SysName disaggregates attention and MoE layers onto separate sets of GPU worker nodes, enabling each layer type to be provisioned and scaled independently. Building on this architecture, \SysName aims to improve MoE efficiency by balancing activated-expert load across GPUs, and support fine-grained scaling to minimize resource costs under SLO constraints. 
However, realizing these goals requires addressing three key challenges in disaggregated MoE inference.


First, disaggregation introduces an ``m-to-n'' communication between $m$ attention and $n$ MoE serving instances at every layer, which significantly increases the end-to-end inference latency.
A naive implementation that adopts pairwise communication would incur many small cross-node transfers across $O(m \times n)$ instance pairs, whose overhead can dominate the end-to-end latency.
\SysName addresses this challenge by trading a modest increase in aggregate data volume for fewer, larger data transfers. 
It employs an \emph{adaptive two-phase transmission scheme} that first aggregates intermediate data from intra-node instances and then performs bulk transfers to the destination nodes. This design significantly reduces the number of data transfers between two sub-clusters and, in turn, reduces overall inference latency.


Second, disaggregation creates a larger and more flexible pool of GPUs for serving expert activations, but also introduces a non-trivial layer-wise scheduling problem---determining, at inference time, how expert activation requests for each layer should be distributed across GPUs to balance load and minimize inference latency. 
This decision must be made with extremely low overhead because layer-wise MoE computation typically completes within a few hundred microseconds (see Fig.~\ref{fig:activation_pattern}). 
To address this challenge, \SysName introduces the \emph{activated-expert-balanced scheduling} that uses a fast heuristic to minimize the number of activated experts per MoE instance. The scheduler runs as a GPU kernel, avoids CPU--GPU synchronization, and operates in a fully distributed manner without cross-GPU coordination.
Consequently, \SysName sustains the scheduling and synchronization overhead at the microsecond level, effectively satisfying the tight latency requirements for layer-wise MoE execution.

The third challenge is to maximize overall resource efficiency, measured by throughput per GPU, while meeting SLO constraints. This necessitates carefully determining resource allocation for both attention and MoE layer types and deciding how experts are placed across MoE instances. 
\SysName addresses this challenge with a \emph{fine-grained, SLO-aware resource scaling scheme} that incrementally adjusts the number of attention and MoE instances to meet token-level SLOs with minimal GPU resources. 
This scheme also optimizes expert placement to minimize the expected number of co-activated experts on each MoE instance. 
Together with the activation scheduler, this design reduces inference latency and improves SLO attainment while lowering resource cost.

We implement \SysName on top of SGLang~\cite{sglang_git} and evaluate it on representative MoE models including DeepSeek-V2~\cite{deepseekv2} and Qwen3-235B~\cite{qwen3}. 
Experiments show that \SysName improves per-GPU throughput by up to 4.7$\times$ and 3.3$\times$ over state-of-the-art monolithic and disaggregated MoE inference systems, respectively, while meeting SLO requirements. We further show that, under real-world LLM inference traces, \SysName adapts to changing workloads by adjusting attention-side and MoE-side configurations at fine granularity, reducing resource cost by about 40\% compared with baselines. 
These results demonstrate that \SysName effectively achieves token-level SLO attainment at low resource cost.

\section{Background and Motivation}
\label{sec:motivation}





\subsection{MoE Inference}
\label{sec:background}

\begin{table}[t]
\centering
\caption{Memory footprint of state-of-the-art MoE models.}
\label{tab:model_memory}
\resizebox{\linewidth}{!}{ 
\small
\begin{tabular}{lccc}
\toprule
\textbf{Model} & \textbf{Expert Mem. (GB)} & \textbf{Total Mem. (GB)} & \textbf{Ratio (\%)} \\
\midrule
Qwen3-235B \cite{qwen3} & 423.0   & 438   & 96.5 \\
DS-V2 \cite{deepseekv2}   & 421.0  & 472.0  & 89.2 \\
DS-V3/R1 \cite{deepseekv3, deepseekr1}   & 1258.0 & 1342.0 & 93.7 \\
Grok-1 \cite{grok1}       & 586.0  & 628.0  & 91.7 \\
\bottomrule
\end{tabular}
}
\end{table}


Mixture-of-Experts (MoE) has become a widely adopted model architecture due to its advantages in scaling LLM capacities~\cite{deepseekv3,xdeepserve,qwen3}. 
Modern MoE models have grown rapidly in size, with expert parameters accounting for most of the model memory footprint (Table~\ref{tab:model_memory}). For example, DeepSeek-V3 contains 256 experts in each MoE layer, and its expert parameters alone account for 93.7\% of the total model memory footprint~\cite{deepseekv3}. 
Serving such models therefore requires substantial GPU memory capacity---fully loading DeepSeek-V3 requires at least 16 H100 GPUs. At the same time, online inference workloads exhibit highly dynamic request arrivals~\cite{blitzscale,yu2025lambdascaleenablingfastscaling,SpotServe}, making static provisioning inefficient: insufficient capacity leads to token-level SLO violations such as high TPOT, whereas overprovisioning for peak load leaves expensive GPUs under-utilized during low-demand periods. 
Consequently, scalable MoE inference systems must elastically adapt to fluctuating workloads while satisfying stringent SLOs at low resource cost~\cite{lina,moelighting,megascale,pregatedmoe}.


LLM inference consists of two phases: prefill and decode. Prefill processes the input prompt in a single forward pass and generates the first token. 
Decode then autoregressively outputs the subsequent tokens, one per iteration. This paper primarily focuses on scalable MoE inference for the decode phase. 
Compared with prefill, decoding typically dominates user-perceptible latency as its per-token cost accumulates over long generation sequences. This decode-centric focus is also well aligned with the emerging deployment practice. 
Recent systems increasingly separate prefill and decode phases~\cite{DistServe,splitwise}, and deployments such as Prefill-as-a-Service~\cite{prfaas} offload long-context, compute-intensive prefill traffic to dedicated remote pools. 
Therefore, local clusters are often left to primarily serve decode traffic together with light prefill requests, a setting that closely matches the scenario targeted by \SysName.\footnote{Our work is complementary to a broader prefill/decode disaggregation, which we discuss in \S\ref{sec:discussion}.}

\subsection{Characteristics and Requirements of MoE Inference}
\label{sec:requirements}


In this section, we characterize the decode-phase MoE inference and derive three key requirements (\emph{\textbf{R1}}--\emph{\textbf{R3}}). 


\begin{figure}[t]
    \centering
    \includegraphics[width=0.95\linewidth]{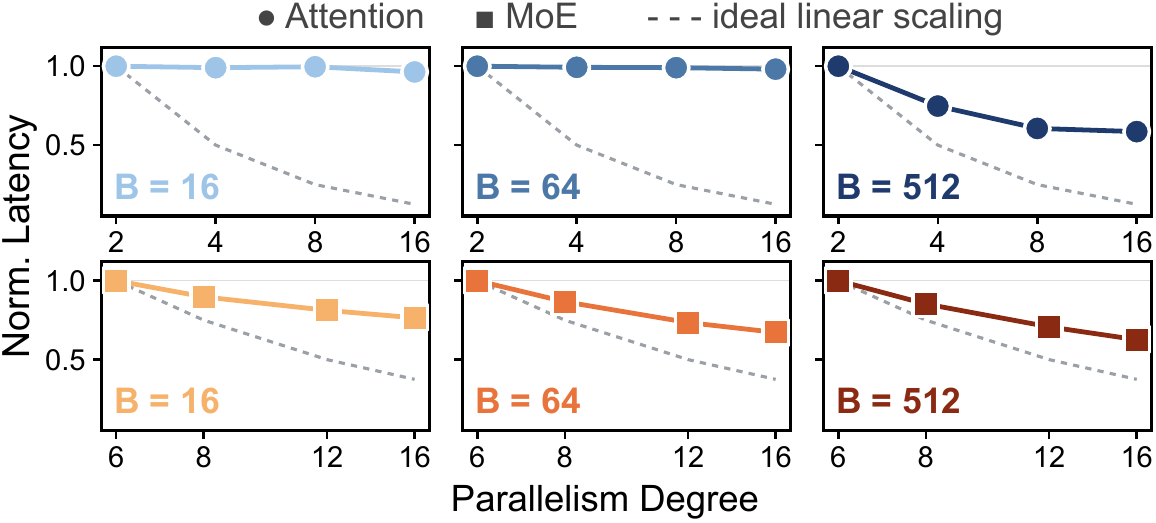}
    \caption{Latency of DeepSeek-V2 attention and MoE layers under different parallelism degrees and batch sizes. Each panel is normalized to the latency at the smallest parallelism degree, and the dashed line shows ideal linear scaling.}
    \label{fig:attn_scaling}
    \vspace{-.2in}
\end{figure}

\subtitle{Distinct patterns of attention and MoE layers.}
We first study the scaling behavior of attention and MoE layers.
Fig.~\ref{fig:attn_scaling} reports the normalized latency of DeepSeek-V2 attention and MoE layers as we vary the parallelism degree under different batch sizes.
The two layer types exhibit markedly different scaling patterns.
For attention layers, increasing the parallelism degree provides little latency benefit at small and moderate batch sizes ($B=16$ and $B=64$); latency only decreases noticeably at a large batch size ($B=512$).
In contrast, MoE layers benefit more consistently from larger MoE-side parallelism across all evaluated batch sizes, although the speedup still remains sublinear.
These results show that a single shared parallelism degree is inefficient for MoE inference.

\begin{figure}[t]
    \centering
    \includegraphics[width=1\linewidth]{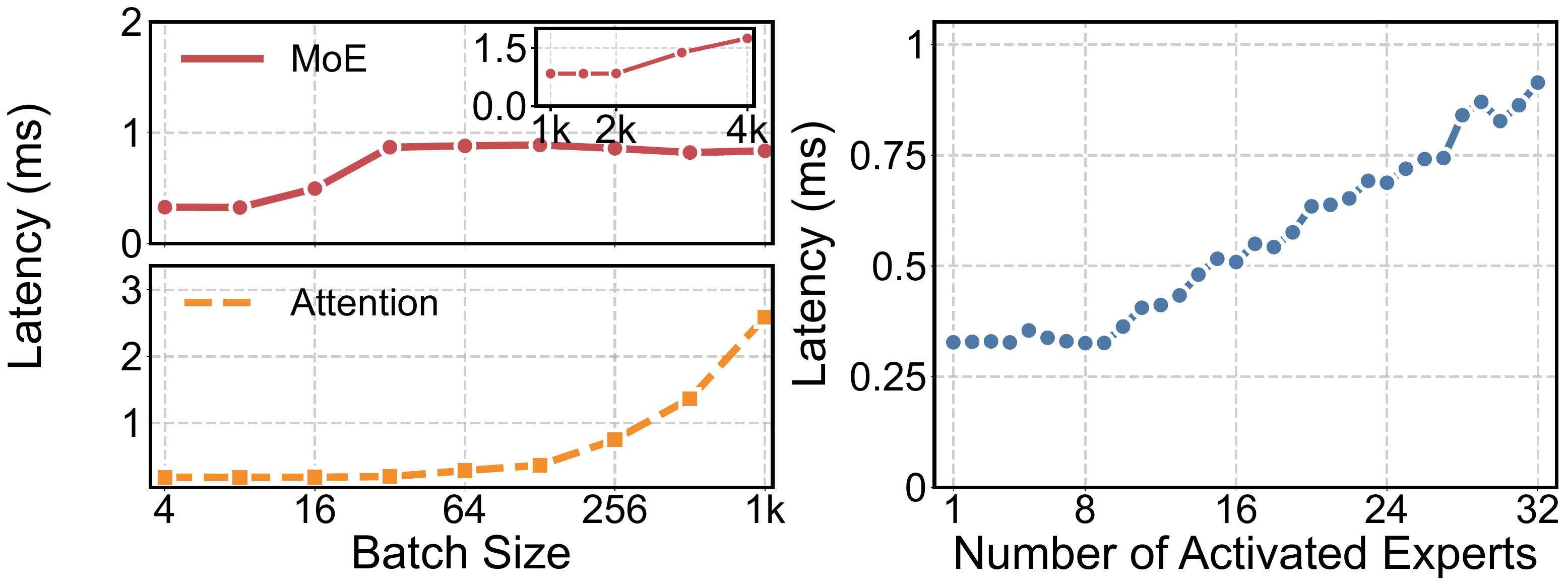}
    \caption{Performance of attention and MoE layers of DeepSeek-V2. Left: latency comparison between attention and MoE layers under different batch sizes. Right: latency of an MoE layer under different numbers of activated experts.}
    \label{fig:moe_attn}
    \vspace{-.2in}
\end{figure}

We further analyze the latency patterns of attention and MoE layers, and highlight their differences.
Using DeepSeek-V2~\cite{deepseekv2} as a representative MoE model, we measure both layers' latency on a single H100 GPU while varying the batch size.
For the attention layer, we fix the input sequence length to 512, following prior work~\cite{step3}.
For the MoE layer, the GPU hosts 32 experts, and each token activates one expert under the balanced top-1 routing.
Fig.~\ref{fig:moe_attn} (left) shows that attention and MoE layers scale differently with batch size. Attention latency remains low at small and moderate batch sizes, but rises sharply once the batch size exceeds 256. MoE latency, in contrast, increases at small batch sizes and then remains nearly flat until the batch size reaches thousands, a regime rarely seen in online decode serving. This difference illustrates that these two layer types do not benefit from the same scaling strategy, as observed in Fig.~\ref{fig:attn_scaling}.


\reqbox{R1. Independent provisioning}{Attention and MoE layers have distinct demands and latency patterns, requiring independent resource provisioning for each layer type.}

\subtitle{Performance bottleneck of MoE Layers.}
We next examine the performance bottleneck of MoE layers, which dominate the model resource footprint with expert parameters accounting for over 90\% of memory usage in modern MoE models (Table~\ref{tab:model_memory}). We begin with a roofline analysis~\cite{roolfline,newllmbottle,moelighting}. In an MoE layer, each expert is dominated by two General Matrix Multiplication (GEMM) operations. Let $d_h, d_e$ denote the hidden and expert intermediate dimensions, respectively. For an expert with batch size $b$, its arithmetic intensity is approximately $I_e \approx 2bd_hd_e / 2d_ed_h = b$. To operate in the compute-bound regime, $I_e$ must exceed $\pi/\beta$, where $\pi$ is the peak FLOPs and $\beta$ is the memory bandwidth of the target hardware.

Consider an MoE layer with $n$ experts under top-$k$ uniform routing. The expected per-expert batch size is $b = B \cdot k/n$, where $B$ is the layer-wise batch size. Therefore, the minimum $B$ required to reach the compute-bound regime is $B \ge \pi n / \beta k$. For example, H100 and A100 GPUs provide 989 and 312 TFLOPs/s of peak compute, and 3.35 and 2.0 TB/s of memory bandwidth, respectively. 
Under this roofline, DeepSeek-V3 would require a layer-wise batch size of about 18k tokens on H100 and 5k tokens on A100 GPUs to become compute-bound, far above typical online inference settings where per-model-instance batch sizes are often below 100~\cite{splitwise,llmtrace2024,polyserve}.

\begin{figure}[t]
    \centering
    \includegraphics[width=0.95\linewidth]{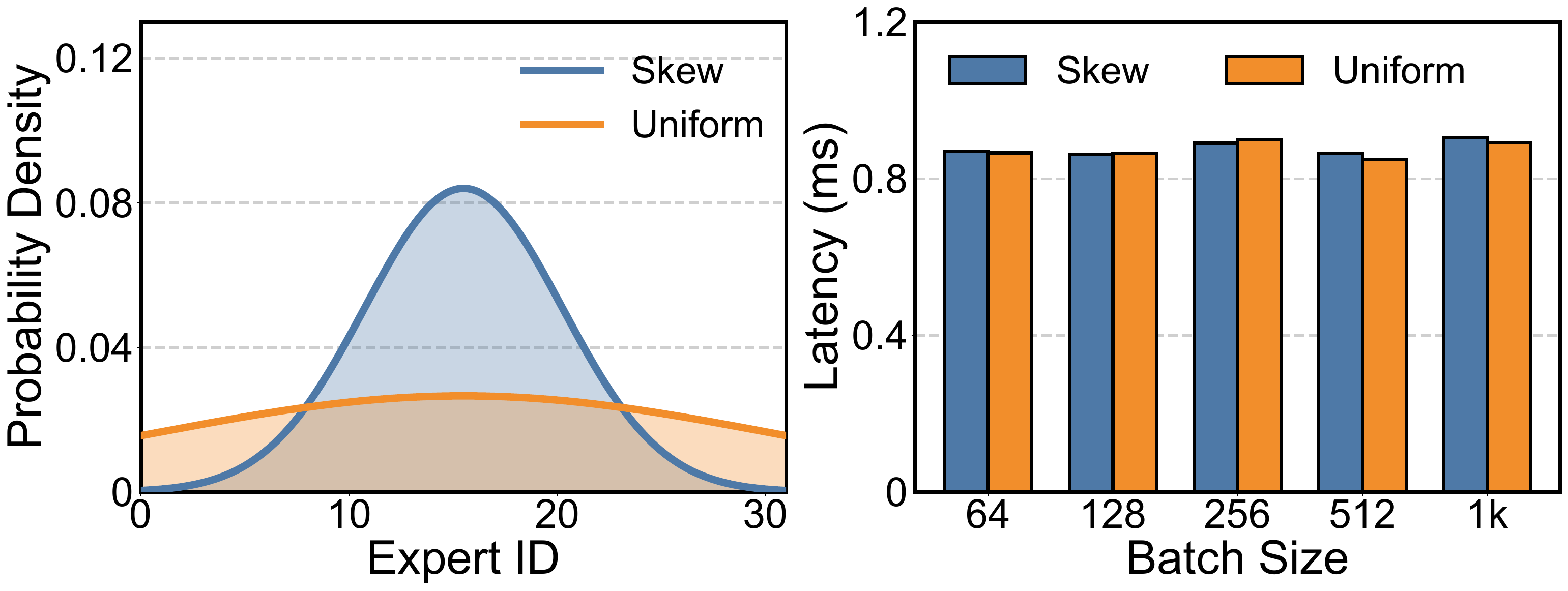}
    \caption{Two distributions of expert activation (left) and latency of an MoE layer under the activation patterns (right).}
    \label{fig:activation_pattern}
    \vspace{-.15in}
\end{figure}

We validate this analysis using a 32-expert MoE layer from DeepSeek-V2~\cite{deepseekv2} on one H100 GPU, emulating the expert density of full-model serving.
Fig.~\ref{fig:moe_attn} (right) isolates the effect of activated expert count by fixing the batch size to 64 and varying the number of activated experts. 
The latency increases approximately linearly with the number of activated experts; only when very few experts are activated does the near-constant kernel-launch overhead dominate. 
This result shows that, in the online decoding regime, MoE latency is primarily determined by the number of distinct activated experts.
We further examine whether the token volume or activation skew changes this conclusion. 
Fig.~\ref{fig:activation_pattern} reports the latency of the same 32-expert MoE layer under different batch sizes and activation distributions. 
For each batch size, all 32 experts are activated at least once, while the activation distribution varies from uniform to skewed. 
The results show that increasing the batch size has only a marginal impact on latency, and that the uniform and skewed activation patterns lead to nearly identical latency. 
Together, our roofline analysis and measurements show that MoE layers are memory-bound and their latency depends strongly on the number of distinct experts to activate.


This property becomes more significant in disaggregated MoE inference when experts are distributed across a larger number of GPUs. 
Let $a_i$ denote the number of activated experts on MoE instance $i$ for a layer, and $a_{\max}=\max_i a_i$.
Since the layer cannot finish until the slowest instance completes, MoE latency is determined by the instance with the largest activated-expert count. 
Therefore, balancing token counts or routing probabilities alone is insufficient; the system should instead balance the activated-expert counts across GPUs.

\reqbox{R2. Activated-expert balancing}{In typical online workloads, MoE layers are largely memory-bound and their latency is dominated by the number of distinct activated experts. Thus, the system should balance the activated-expert counts across GPUs to minimize the MoE execution latency.}

\begin{figure}[t]
    \centering
    \includegraphics[width=1\linewidth]{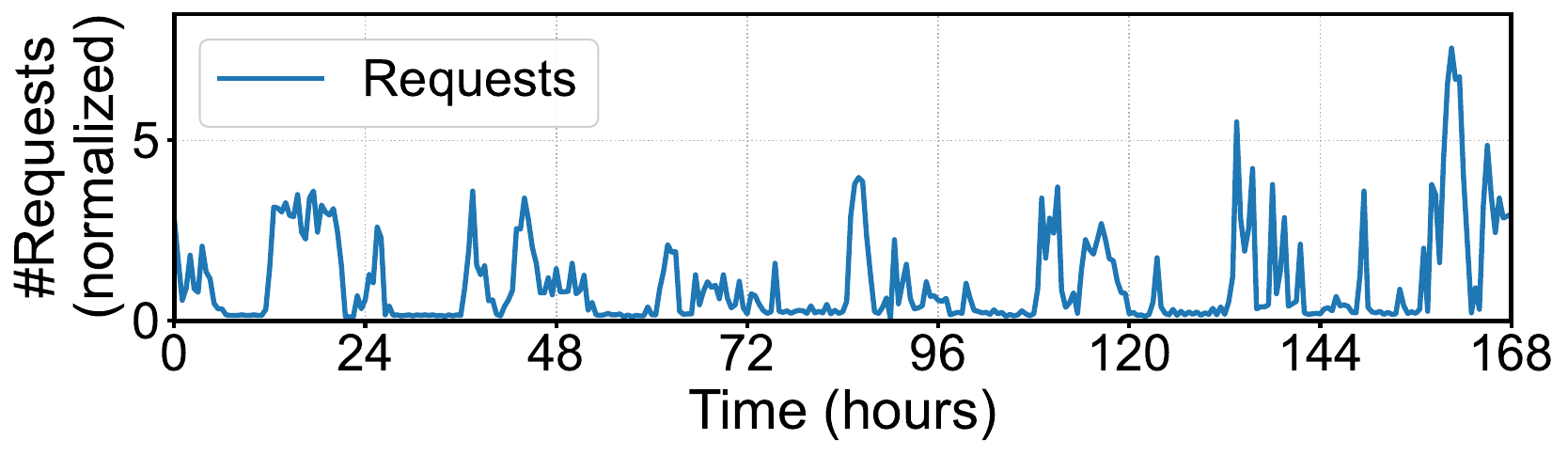}
    \caption{One-week production LLM serving trace (normalized to the trace-wide mean). Request arrival rate is highly bursty, with peaks reaching $\sim$7.5$\times$ the mean.}
    \label{fig:bursty_trace}
    \vspace{-.2in}
\end{figure}

\subtitle{Dynamic workloads and scaling requirements.}
Online LLM serving workloads are highly dynamic. 
Fig.~\ref{fig:bursty_trace} shows a one-week production trace whose request arrival rate exhibits clear diurnal patterns and reaches around 7.5$\times$ of the trace-wide mean. 
Such burstiness makes static provisioning inefficient: provisioning for the average load risks token-level SLO violations during bursts, whereas provisioning for the peak keeps expensive GPUs under-utilized during low-demand periods.
More importantly, workload changes also shift the optimal resource configurations of MoE inference. 
As shown in Fig.~\ref{fig:attn_scaling}, different batch sizes favor different parallelism degrees and lead to different latency patterns across attention and MoE execution. 
Therefore, the system should resize attention- and MoE-side resources adaptively as workload changes, ensuring that token-level SLOs are satisfied at low resource cost.

\reqbox{R3. Fine-grained elasticity under SLOs}{No single resource configuration remains efficient under dynamic workloads. The system should therefore adjust attention- and MoE-side resources in a fine-grained and incremental manner to satisfy token-level SLOs at low resource cost.}

\subsection{Limitations of Existing Solutions}
\label{sec:characteristics} 

Existing MoE inference systems can be broadly categorized into two classes, \textit{monolithic} and \textit{disaggregated systems}, as summarized in Table~\ref{tab:comparison}. However, neither of them satisfies the three key requirements outlined in \S\ref{sec:requirements}.

\begin{table}[t]
\centering
\caption{Comparison of existing MoE inference systems.}
\label{tab:comparison}
\resizebox{\columnwidth}{!}{
\begin{tabular}{lccc}
\toprule
\begin{tabular}[c]{@{}l@{}}\textbf{System}\end{tabular} &
\begin{tabular}[c]{@{}c@{}}\textbf{Independent}\\\textbf{Provisioning}\end{tabular} &
\begin{tabular}[c]{@{}c@{}}\textbf{Activated-Expert}\\\textbf{Balancing}\end{tabular} &
\begin{tabular}[c]{@{}c@{}}\textbf{Fine-grained}\\\textbf{Elasticity}\end{tabular} \\
\midrule
Monolithic~\cite{lina,vllm_git,sglang_git} & $\times$ & $\times$ & $\times$ \\
MegaScale-Infer~\cite{megascale}           & $\checkmark$ & $\times$ & $*$ \\
xDeepServe~\cite{xdeepserve}               & $\checkmark$ & $\times$ & $\times$ \\
EaaS~\cite{eaas}                           & $\checkmark$ & $\times$ & $\times$ \\
\textbf{\SysName}                          & $\checkmark$ & $\checkmark$ & $\checkmark$ \\
\bottomrule
\end{tabular}}
\vspace{-.2in}
\end{table}

\subtitle{Monolithic MoE inference.} 
Most MoE inference systems adopt a monolithic design. 
Systems such as SGLang~\cite{sglang_git}, vLLM~\cite{vllm_git}, and LINA~\cite{lina} co-locate attention and MoE layers on the same GPUs, use a shared parallelism configuration, and scale by replicating or reconfiguring full model instances. 
This design falls short of all three requirements. 
First, monolothic configurations cannot match the distinct latency patterns and demands of attention and MoE layers.
It also couples their memory requirements: MoE expert parameters and attention-side KV caches  share the same GPU memory budget, forcing the system to over-provision for the combined peak footprint (\emph{\textbf{R1}}).
Second, monolithic systems expose little control over layer-wise expert activation scheduling, and thus cannot balance activated-expert counts across GPUs to reduce MoE latency (\emph{\textbf{R2}}). 
Third, their elasticity is inherently coarse-grained: the smallest scaling unit is a full model replica---for example, at least 16 H100 GPUs for DeepSeek-V3 (Table~\ref{tab:model_memory})---and scaling requires loading all parameters and rebuilding parallelism groups. 
Such coarse reconfiguration cannot efficiently track dynamic workloads under SLO constraints (\emph{\textbf{R3}}).

\subtitle{Disaggregated MoE inference.}
Beyond monolithic designs, recent systems separate attention and MoE execution onto distinct nodes~\cite{megascale,xdeepserve,eaas}. 
By decoupling the two layer types, these systems partially address independent provisioning (\emph{\textbf{R1}}). 
However, this architectural separation alone is insufficient, and existing solutions still fall short of the remaining requirements.
First, they generally overlook the need to balance activated-expert counts across GPUs (\emph{\textbf{R2}}).
Their MoE-side mechanisms mainly focus on hot-expert replication or token balancing, such as evenly distributing tokens across expert replicas.
While such strategies can reduce token imbalance, they do not directly minimize $a_{\max}$, the maximum number of distinct activated experts across GPUs, which determines MoE latency as discussed in \S\ref{sec:requirements}.
As a result, even a token-balanced configuration can leave one GPU activating more distinct experts than others, making it the straggler that limits latency improvement (see \S\ref{sec:evaluation} and Fig.~\ref{fig:ep_latency_box}).

Second, their elasticity remains limited under dynamic workloads (\emph{\textbf{R3}}).
EaaS~\cite{eaas} enables flexible reconstruction of communication channels between attention and MoE instances, and xDeepServe~\cite{xdeepserve} provides specialized support for attention--expert disaggregation on NPU superpods. 
These mechanisms improve the data plane for disaggregated execution, but they do not target resource configurations of the attention and MoE sides under changing workloads.  
MegaScale-Infer~\cite{megascale} provides partial support by tuning the attention-to-MoE resource ratio to balance the execution times of the two sides; however, this design restricts the feasible configuration space and leads to suboptimal performance or even SLO violations (see \S\ref{sec:e2e} and Fig.~\ref{fig:dsv2_e2e_scale}). 
Consequently, existing disaggregated systems still lack the fine-grained elasticity needed to track workload changes at low resource cost.

\section{\SysName Design}
\label{sec:system}



In this section, we present \SysName, a scalable and resource-efficient MoE inference system.

\subsection{Design Principles and Challenges}
\label{sec:insights}

Guided by the three requirements in \S\ref{sec:requirements}, \SysName adopts three design principles, each introducing a key challenge.

\emph{First, \SysName needs to disaggregate attention and MoE layers onto separate sub-clusters to enable module-specific resource allocation and scaling} (\emph{\textbf{R1}}).
This design creates the flexibility needed to provision the two sides independently, but it also moves layer-wise activation transfer across sub-clusters.
At every MoE layer, $m$ attention instances must exchange activations with $n$ MoE instances.
Naively issuing these $m$-to-$n$ transfers creates many small messages on the inference's critical path, making communication overhead the first challenge.

\emph{Second, \SysName needs to balance the activated-expert counts across GPUs through layer-wise activation scheduling} (\emph{\textbf{R2}}).
To reduce MoE latency, the scheduler must route activation requests to experts to minimize the maximum number of distinct activated experts across GPUs, $a_{\max}$. 
However, this decision is made at every MoE layer and every decoding step, where MoE computation may finish within only hundreds of microseconds.
The second challenge is therefore to achieve activated-expert balancing with a microsecond-level scheduling overhead and without expensive global coordination.

\emph{Third, \SysName needs to optimize resource efficiency, measured by per-GPU throughput, while satisfying token-level SLOs} (\emph{\textbf{R3}}).
This requires jointly determining resource allocations for attention and MoE instances, and the placement of expert replicas.
The third challenge is to find and apply the SLO-feasible configurations dynamically as workloads change, while keeping resource cost low. 

\subsection{Design Overview}
\label{sec:sys_overview}


\begin{figure}[t]
    \centering
    \includegraphics[width=0.92\linewidth]{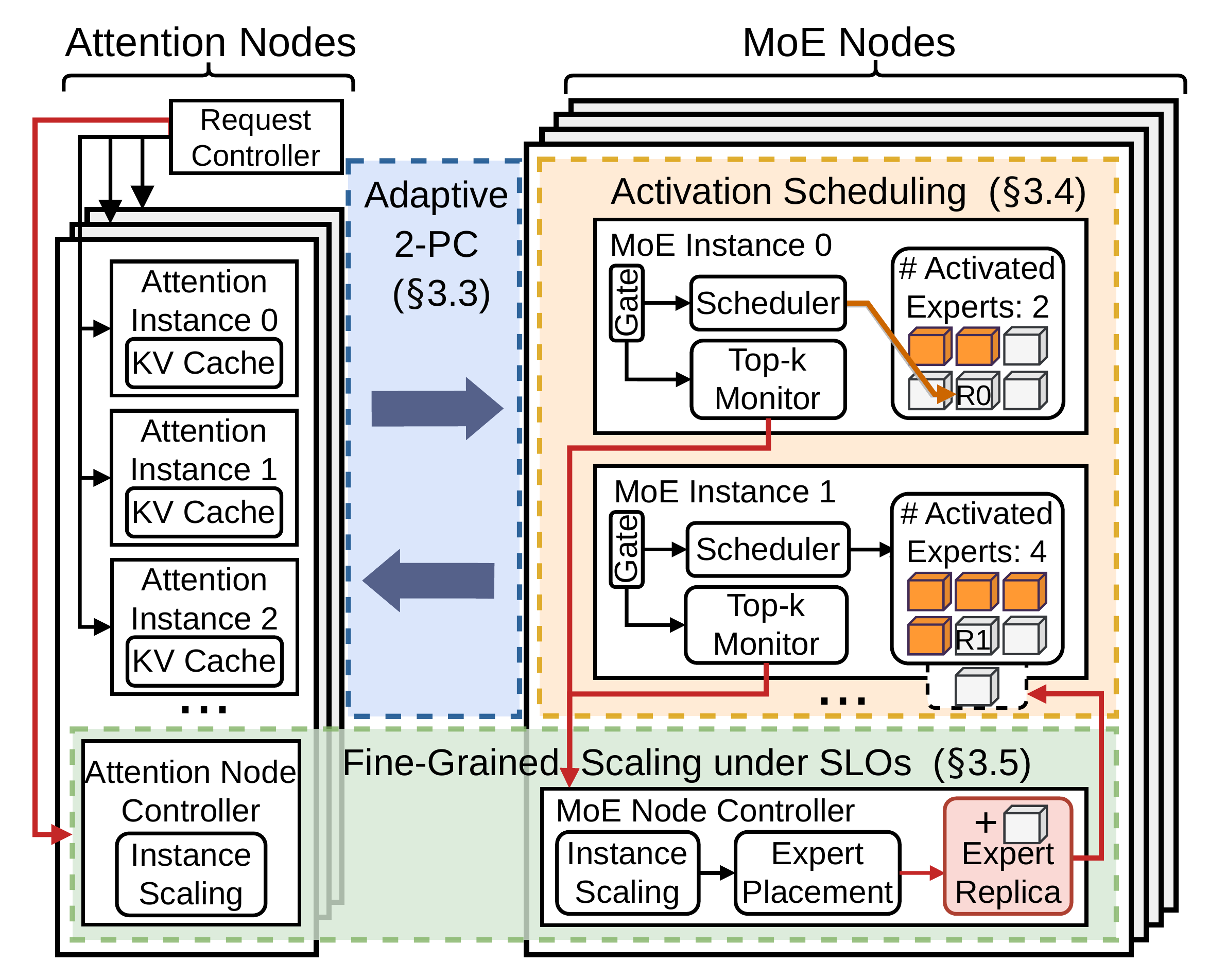}
    \caption{Architecture overview of \SysName.}
    \label{fig:arch}
    \vspace{-.2in}
\end{figure}

\begin{figure*}[!t]
    \centering  
    \includegraphics[width=0.92\linewidth]{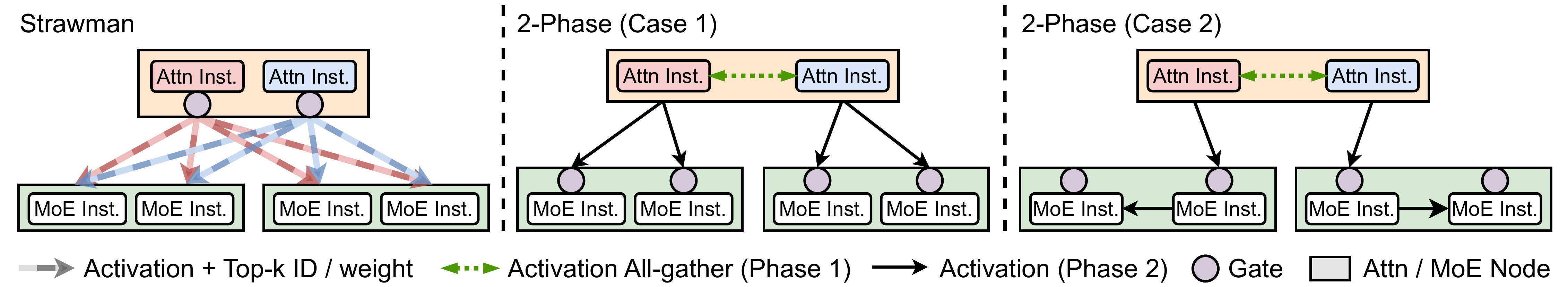}
    \caption{Comparison between a strawman solution (left) and adaptive two-phase communication (middle and right).}
    \label{fig:m2n}
    \vspace{-.2in}
\end{figure*}

Fig.~\ref{fig:arch} shows the architectural overview of \SysName. The cluster consists of two sub-clusters of GPU nodes, attention nodes and MoE nodes.  These nodes  are independently managed and scaled within the two sub-clusters.
\SysName supports low-latency data transfers between sub-clusters with our adaptive two-phase communication mechanism (\S\ref{sec:sys_2pc}).

On the attention side, each attention node hosts multiple attention instances. 
Each instance runs on one GPU and keeps a full replica of the attention layers.\footnote{\SysName primarily targets state-of-the-art MoE models with many small-to-moderate-sized experts, where data parallelism and expert parallelism dominate~\cite{deepseekv3}.}
There is a request controller which assigns incoming requests to attention instances. 

On the MoE side, each MoE node hosts multiple MoE instances, each running on one GPU and storing a subset of expert replicas. 
To achieve the activated-expert balancing, each MoE instance runs a lightweight activation scheduler that maps the expert activation requests to expert replicas for minimizing the maximum number of activated experts across GPUs. 
The scheduler is implemented as a GPU kernel and runs in a distributed, synchronization-free manner (\S\ref{sec:lb_scheduling}).
MoE instances also collect activation statistics and report them to the associated MoE controller on the MoE node. 
Together with the attention controller, they periodically adjust attention resources, MoE resources, and expert placement to improve per-GPU throughput under token-level SLOs (\S\ref{sec:exper_mana}).




\subsection{Adaptive Two-Phase Communication}
\label{sec:sys_2pc}

Disaggregating attention and MoE layers turns intra-instance data movement into cross-sub-cluster communication.
With $m$ attention instances and $n$ MoE instances, each MoE layer requires dispatching activations from attention instances to MoE instances and aggregating the results back.
A straightforward implementation lets every attention instance directly communicate with every MoE instance, as shown in Fig.~\ref{fig:m2n} (left).
This design incurs $O(m\times n)$ point-to-point transfers and creates many small messages on the inference critical path.
Additionally, existing collective communication mechanisms are not well suited for this setting.
Cluster-wide collectives such as NCCL~\cite{nccl} and MSCCL++~\cite{msccl++}, along with expert-parallel libraries such as DeepEP~\cite{deepep_git}, are mainly designed for symmetric communication groups, where all participants follow the same communication pattern.
In contrast, the attention-MoE communication is asymmetric: the two sub-clusters can have different numbers of instances, and their sizes may change as \SysName elastically scales resources.

\SysName therefore designs a customized communication mechanism for the disaggregated MoE inference. 
We observe that each individual transfer between attention and MoE nodes is small, while the large number of transfers dominates the communication overhead. 
Thus, \SysName prioritizes reducing the number of cross-node transfers rather than minimizing the aggregate data volume.

\subtitle{Gating on the MoE side.}
\SysName places the gating network on the MoE side to simplify communication. 
As shown in Fig.~\ref{fig:m2n} (left), a strawman design is to place the gate on the attention side and transmit only the activations routed to each expert. 
Although this reduces the total amount of activation data, it requires sending routing metadata together with activations and reorganizing activation tensors according to expert destinations. 
This either increases the number of small transfers or introduces extra packing and memory re-layout overheads. 
Since our setting is dominated by small-transfer overhead, such fine-grained dispatch is inefficient. 
\SysName instead sends complete activations to the MoE side and performs gating there, which reduces the communication complexity and avoids per-expert tensor packing on attention nodes.

\subtitle{Adaptive two-phase communication.}
\SysName further reduces communication overhead with a two-phase scheme that leverages fast intra-node communication before inter-node transfer. 
In the first phase, multiple instances on the same source node locally aggregate intermediate activations through NVLink-based collective primitives. This aggregation produces larger cross-node payloads. In the second phase, these aggregated payloads are sent to the destination nodes.

\SysName adaptively selects between two transfer regimes, as shown in Fig.~\ref{fig:m2n} (middle and right). 
\textbf{Case-1:} When each attention node only needs to send data to a small number of MoE nodes, the aggregated payloads are directly transmitted to the corresponding destination nodes. 
\textbf{Case-2:} When the number of destinations or the data volume is large, \SysName uses a one-to-one inter-node transmission pattern. 
Each attention node sends aggregated activations to a designated MoE node, which then distributes the data to local MoE instances via an intra-node NVLink multicast. 
\SysName adaptively selects between these two regimes based on resource configuration and traffic load.
Communication in the reverse direction (i.e., MoE to attention) follows the same two-phase principle, using intra-node all-reduce to aggregate intermediate results on the MoE side before sending them to attention nodes.

\subsection{Activated-Expert-Balanced Scheduling}
\label{sec:lb_scheduling}

\begin{figure}[t]
    \centering
    \includegraphics[width=.92\linewidth]{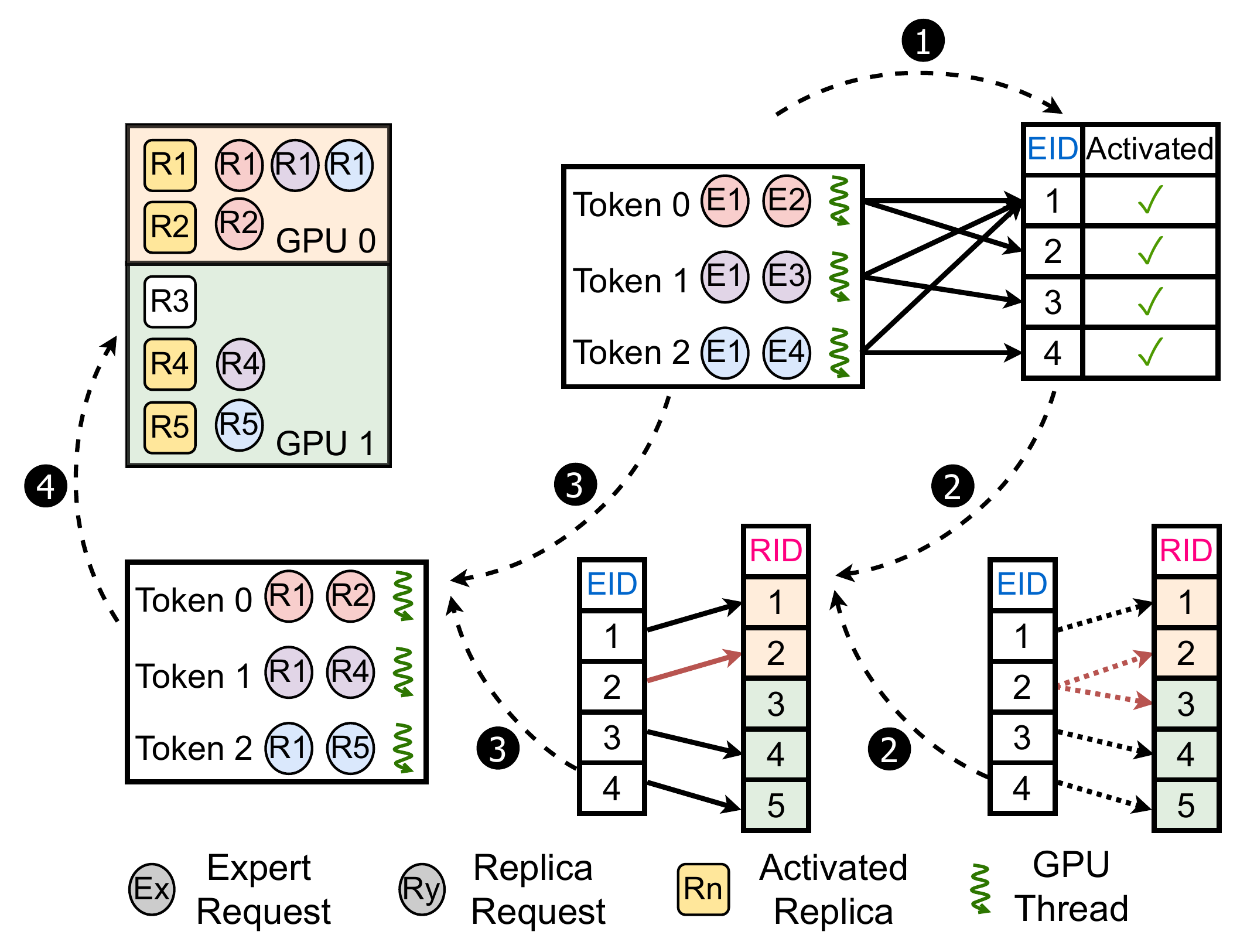}
    \caption{Scheduling workflow of \SysName.}
    \label{fig:scheduler}
    \vspace{-.2in}
\end{figure}

As established in \S\ref{sec:requirements}, the MoE layer latency is determined by the number of activated experts on the bottlenecked MoE instance. \SysName therefore needs to schedule the expert activation requests across expert replicas so as to balance the activated-expert counts at every MoE layer. 
This scheduling is challenging for two reasons. 
First, finding the optimal assignment is a combinatorial load-balancing problem over all possible mappings from activated experts to replicas, making it prohibitively expensive to solve online for every layer. 
Second, making such decisions requires fine-grained activation information, such as top-$k$ routing results, and therefore introduces frequent CPU-GPU synchronization or cross-GPU coordination. 
The resulting overhead can be substantial and may easily exceed the MoE execution time itself, which is often only a few hundred microseconds.

\begin{algorithm}[t]
\caption{Activated-Expert-Balanced Scheduling}
\label{algo:aebs}
\small
\textbf{Input:} \\
-- $T$: number of tokens, $n_e$: number of MoE instances \\
-- $k$: number of activated experts per token \\
-- $L(i,j)$: logical expert ID of the $j$-th activated expert for token $i$ \\
-- $R(e)$: number of replicas for expert $e$ \\
-- $\mathcal{G}(e)$: set of instances hosting replicas of expert $e$ \\
-- $P(e,g)$: physical replica ID of expert $e$ on instance $g$ \\
\textbf{Output:} \\
-- $O(i,j)$: physical replica ID of the $j$-th activated expert for token $i$

\begin{algorithmic}[1]

\State $\mathcal{E} \gets \bigcup_{i=1}^{T} \bigcup_{j=1}^{k} \{ L(i,j) \}$ \Comment{Collect all activated experts}
\State Initialize $\mathrm{actRep}[e] \gets -1$ for all $e \in \mathcal{E}$
\State Initialize $\mathrm{load}[g] \gets 0$ for all $g \in \{1,2,\dots,n_e\}$

\Statex \textit{\# Assign multi-replica experts via load balancing}
\ForAll{$e \in \mathcal{E}$ where $R(e) = 1$}
    \State $g \gets$ the unique instance in $\mathcal{G}(e)$
    \State $\mathrm{actRep}[e] \gets P(e,g)$
    \State $\mathrm{load}[g] \gets \mathrm{load}[g] + 1$
\EndFor

\Statex \textit{\# Assign multi-replica experts via load balancing}
\ForAll{$e \in \mathcal{E}$ where $R(e) > 1$}
    \State $g^* \gets \arg\min_{g \in \mathcal{G}(e)} \mathrm{load}[g]$
    \State $\mathrm{actRep}[e] \gets P(e,g^*)$
    \State $\mathrm{load}[g^*] \gets \mathrm{load}[g^*] + 1$
\EndFor

\Statex \textit{\# Map tokens' activation requests to physical replicas}
\For{$i = 1$ to $T$}
    \For{$j = 1$ to $k$}
        \State $O(i,j) \gets \mathrm{actRep}[L(i,j)]$
    \EndFor
\EndFor

\end{algorithmic}
\end{algorithm}

\subtitle{Scheduling workflow.}
\SysName introduces a lightweight activation scheduling workflow, as shown in Fig.~\ref{fig:scheduler}.
For each MoE layer, MoE-side gating first produces the top-$k$ logical expert IDs (EIDs) for all tokens in the current decode batch.
\SysName then scans these routing results and collects the union of selected EIDs, i.e., the set of activated logical experts in this batch (Step 1).
This step is implemented as a GPU kernel, with tokens processed in parallel by GPU threads.
Given the activated logical experts and the expert-replica mapping, \SysName selects one physical replica ID (RID) for each activated EID (Step 2).
For replicated experts, \SysName chooses the replica on the currently least-loaded MoE instance, where load is measured by the number of activated experts assigned to that instance in the current layer.
After replica selection, \SysName rewrites each token's routing result from logical EIDs to the selected RIDs (Step 3), and dispatches token activations to the MoE instances that host those replicas (Step 4).
In the example in Fig.~\ref{fig:scheduler}, \SysName selects replicas that balance activated-expert counts across GPUs, rather than merely balancing token counts.

\subtitle{Scheduling algorithm.} 
\SysName implements the aforementioned workflow with an Activated-Expert-Balanced Scheduling (AEBS) algorithm. 
AEBS greedily reduces the maximum number of activated experts on any MoE instance (Algorithm~\ref{algo:aebs}). 
It first collects the set of experts activated by the current batch (line 1). 
It then assigns single-replica experts to their unique hosting instances and schedules multi-replica experts to the least-loaded instances among those hosting their replicas (lines 2-11). 
This yields a near-balanced expert activation across instances while incurring only negligible computational overhead.

\subtitle{Synchronization-free scheduling.}
To avoid the overhead of global coordination, \SysName makes AEBS synchronization-free across MoE instances via two mechanisms.
First, \emph{\SysName implements AEBS as a GPU kernel to achieve microsecond-level scheduling latency}. 
This avoids CPU--GPU synchronization when accessing the per-token top-$k$ routing results, and allows many tokens to be processed in parallel (i.e., steps 1 and 3 in Fig.~\ref{fig:scheduler}). 
Second, \emph{\SysName trades a small amount of redundant computation to eliminate cross-instance synchronization}. 
Instead of using a centralized global scheduler, each MoE instance independently runs the same AEBS kernel with identical input, including token activation patterns, replica layout, and instance metadata. 
Since AEBS is deterministic with respect to these inputs, all instances compute the same global assignment from logical experts to physical replicas. 
\SysName updates metadata such as replica layout only when the MoE sub-cluster is reconfigured, which occurs at a much coarser time scale (e.g., on the order of hours) than per-layer execution, making the propagation overhead negligible (\S\ref{sec:exper_mana}). 
The redundant scheduling computation on each GPU is also small compared with the MoE forward computation. 
As a result, \SysName eliminates inter-GPU communication for activation scheduling while preserving correctness and imposing negligible overhead.

\subsection{Fine-Grained Scaling under SLOs}
\label{sec:exper_mana}


We design a fine-grained, SLO-aware resource scaling scheme that jointly selects attention-side and MoE-side resources.
Let $n_a$ and $n_e$ denote the numbers of active attention and MoE instances, respectively, where each instance runs on one GPU. 
Given a workload demand $\lambda$ and a TPOT SLO, \SysName searches for a configuration $(n_a,n_e)$ that can sustain $\lambda$ while keeping the predicted TPOT within the SLO.
In disaggregated MoE inference, scaling becomes a two-dimensional optimization problem rather than instance-level scaling of the full model.
Among all SLO-feasible configurations, \SysName chooses the one with the smallest GPU count $n_a+n_e$, which equivalently maximizes the throughput per GPU. 

As demand changes, \SysName re-runs this optimization and applies the new configuration incrementally.

\PHM{Performance model.}
For a candidate configuration $(n_a,n_e)$, \SysName estimates TPOT with a layer-wise latency model. 
On the attention side, requests are evenly dispatched across the $n_a$ data-parallel attention instances. Let $B$ denote the in-flight decode batch size, $b=B/n_a$ denote the per-instance local batch size, $S_{\mathrm{ctx}}$ denote the average context length, and $L$ denote the number of layers. 
Following prior LLM serving systems~\cite{agrawal2024vidur,DistServe,splitwise}, \SysName models TPOT as the sum of attention, MoE, and communication costs across layers:
\begin{subequations}
\label{eq:perfmodel}
\begin{align}
\mathrm{TPOT}
&= \sum_{\ell=1}^{L}
\bigl[
T_{\mathrm{attn}}^{(\ell)}
+ T_{\mathrm{moe}}^{(\ell)}
+ T_{\mathrm{comm}}^{(\ell)}
\bigr],
\label{eq:tpot} \\
T_{\mathrm{attn}}^{(\ell)}
&= \max\!\Bigl(
c_a^{(\ell)},\;
\alpha^{(\ell)} b + c_{kv}^{(\ell)} b S_{\mathrm{ctx}}
\Bigr),
\label{eq:tattn} \\
T_{\mathrm{moe}}^{(\ell)}
&= \beta^{(\ell)} \cdot
a_{\max}^{(\ell)}(n_e,B)
+ c_e^{(\ell)} .
\label{eq:tmoe}
\end{align}
\end{subequations}

Here $T_{\mathrm{attn}}^{(\ell)}$, $T_{\mathrm{moe}}^{(\ell)}$, and 
$T_{\mathrm{comm}}^{(\ell)}$ denote the attention, MoE, and communication latencies of layer $\ell$, respectively. 
The attention term $T_{\mathrm{attn}}^{(\ell)}$ follows the roofline model~\cite{williams2009roofline}: $c_a^{(\ell)}$ captures the memory-bound latency plateau that dominates at small workloads, while $\alpha^{(\ell)} b + c_{kv}^{(\ell)} b S_{\mathrm{ctx}}$ captures the cost of computation and KV-cache access. 
The MoE term $T_{\mathrm{moe}}^{(\ell)}$ follows the linear dependence on $a_{\max}^{(\ell)}(n_e,B)$, the maximum number of distinct activated experts across MoE instances under the candidate MoE size and AEBS strategy.
The communication term $T_{\mathrm{comm}}^{(\ell)}$ is obtained from the profiled cost model of the adaptive two-phase communication scheme (\S\ref{sec:sys_2pc}). 
All hardware-dependent coefficients, including $\alpha^{(\ell)}$, $\beta^{(\ell)}$, $c_a^{(\ell)}$, $c_{kv}^{(\ell)}$, $c_e^{(\ell)}$, are obtained through a one-time offline profiling.

\PHM{Problem formulation.}
The in-flight batch size is not an independent decision variable. 
Under the steady-state decode serving, it is determined by Little's Law~\cite{little1961}:
\begin{equation}
\label{eq:fixedpoint}
B^* = \lambda \cdot 
\mathrm{TPOT}(B^*,n_a,n_e,S_{\mathrm{ctx}}).
\end{equation}
Thus, changing the resource configuration $(n_a,n_e)$ changes the TPOT curve and also the steady-state batch size $B^*$.

Each candidate configuration must satisfy the per-GPU memory constraints. 
Let $M$ be the memory budget of a GPU, and let $b^*=B^*/n_a$ denote the steady-state local batch size on each attention instance. 
We use $\mathcal{M}_a(b^*,S_{\mathrm{ctx}})$ to denote the memory usage of an attention instance, including the attention weights, KV cache, and activation buffers. 
On the MoE side, memory usage is dominated by the pinned expert weights: each MoE instance pins at most $C$ expert replicas, which makes the per-GPU memory constraint easy to enforce during placement. 
The resource scaling problem is formulated as:
\begin{equation}
\label{eq:scaling_opt}
\begin{aligned}
\min_{n_a,n_e,B^*} \quad
& n_a+n_e \\
\textrm{s.t.} \quad
& \mathrm{TPOT}(B^*,n_a,n_e,S_{\mathrm{ctx}}) \le \mathrm{SLO}, \\
& \mathcal{M}_a(b^*,S_{\mathrm{ctx}}) \le M, \\
& n_e \cdot C \ge E, \\
& n_a,n_e \in \mathbb{Z}^{+}.
\end{aligned}
\end{equation}
The first constraint enforces the TPOT SLO, and the second constraint enforces the attention-side memory feasibility. 
The third constraint ensures that the MoE sub-cluster has enough expert slots to host all expert replicas. 


\PHM{Scaling solution.}
Solving Eq.~\eqref{eq:scaling_opt} has two challenges. 
First, the TPOT model depends on the maximum number of distinct activated experts $a_{\max}^{(\ell)}(n_e,B)$, which is workload- and scheduling-dependent and thus difficult to capture with a static closed-form model.
Second, for each candidate resource configuration $(n_a,n_e)$, the steady-state batch size $B^*$ is unknown in advance; it must be solved from the fixed-point equation in Eq.~\eqref{eq:fixedpoint} before checking SLO and memory feasibility.

\SysName uses recent activation statistics to build a Monte Carlo estimator $\widehat{a}_{\max}^{(\ell)}(n_e,B)$ of $a_{\max}^{(\ell)}(n_e,B)$.
We formulate the top-$K$ routing as a balls-into-bins problem~\cite{raab1998balls} and derive a theoretical upper bound on $a_{\max}$ in Appendix~\ref{appendix:maxa} (Eq.~\ref{eq:max_a_bound}). 
Building on the analysis, \SysName uses a Monte Carlo approach for the $a_{\max}$ estimation and scaling decisions. 
For each candidate $(n_e,B)$ and each MoE layer $\ell$, it samples $B$ tokens from the recent activation trace, applies the current scheduling strategy, and records the resulting estimate $\widehat{a}_{\max}^{(\ell)}(n_e,B)$.
The resulting lookup table $\widehat{a}_{\max}^{(\ell)}(n_e,B)$ is rebuilt periodically, ensuring that the model is aligned with the current workload.

To solve Eq.~\eqref{eq:fixedpoint}, \SysName performs a bounded one-dimensional search for the steady-state batch size $B^*$ over $[1,B_{\max}]$, where $B_{\max}$ is the maximum batch size allowed by the GPU memory budget.
For a fixed configuration $(n_a,n_e)$, we define the residual $f(B)=B-\lambda\cdot\mathrm{TPOT}(B,n_a,n_e,S_{\mathrm{ctx}})$. 
In our profiled operating range, the residual is monotonic and thus \SysName solves it with a bounded binary search~\cite{brent1973algorithms,press2007numerical}.
\SysName handles two boundary cases explicitly. 
If $f(1)\ge0$, the workload is too light to form a larger steady-state batch, so \SysName sets $B^*=1$.
If $f(B_{\max})<0$, even the largest memory-feasible batch cannot sustain the demand, so \SysName discards the current candidate configuration $(n_a,n_e)$.

\SysName then solves Eq.~\eqref{eq:scaling_opt} by enumerating the candidate configurations over a bounded search space.
Configurations that are clearly infeasible, such as $n_e<n_e^{\min}$, are pruned before evaluation. 
Algorithm~\ref{algo:scaling} gives this scaling procedure.  
For each remaining candidate configuration $(n_a,n_e)$, \SysName first solves Eq.~\eqref{eq:fixedpoint} to obtain $B^*$ (line 3), evaluates TPOT, checks memory feasibility, and selects the feasible configuration with the smallest GPU count (lines 6--10). 
This computation incurs negligible runtime overhead: each TPOT evaluation only requires the constant-time lookups of $\widehat{a}_{\max}^{(\ell)}$, and the search space over $(n_a,n_e)$ is bounded by the cluster size. 
The selected configuration $(n_a^\star,n_e^\star)$ is then applied incrementally by adding or removing attention and MoE instances.

\begin{algorithm}[t]
\caption{Fine-Grained, SLO-Aware Resource Scaling}
\small
\label{algo:scaling}
\textbf{Input:} \\
-- $n_{\max}$: upper bound of instance sizes\\
-- $n_e^{\min}$: lower bound of MoE instance sizes, i.e., $\lceil E/C\rceil$\\
-- $B_{\mathrm{max}}$: upper bound of batch sizes according to GPU memory budget\\
\textbf{Output:} \\
-- $(n_a^*,n_e^*,B^*)$: optimal configuration if feasible

\begin{algorithmic}[1]
\State $\mathrm{opt}\gets\bot$;\quad $J^*\gets\infty$
\For{$(n_a,n_e)\in\{1,\dots,n_{\max}\}\times\{n_e^{\min},\dots,n_{\max}\}$}
    \State $B^*\gets$ batch size in $[1,B_{\max}]$ satisfying Eq.~\eqref{eq:fixedpoint}
    \If{$B^*=\bot$}
        \State \textbf{continue}
    \EndIf
    \State $T\gets\mathrm{TPOT}(B^*,n_a,n_e,S_{\mathrm{ctx}})$
    \If{$T>\mathrm{SLO}$ \textbf{or} $\neg\Call{MemoryFeasible}{B^*,n_a,n_e}$}
        \State \textbf{continue}
    \EndIf
    \If{$n_a+n_e<J^*$}
        \State $\mathrm{opt}\gets(n_a,n_e,B^*)$;\quad $J^*\gets n_a+n_e$
    \EndIf
\EndFor
\State \Return $\mathrm{opt}$

\end{algorithmic}
\end{algorithm}

\subtitle{Expert placement at the MoE side.}
After determining the optimal resource configuration, \SysName allocates and places expert replicas to support the activated-expert-balanced scheduling. 
The key goal is to avoid collocating experts that are frequently activated together, which increases the number of distinct activated experts on the same instance and increases MoE latency.
Accordingly, \SysName processes replicas in descending load order and places each replica on the instance that incurs the smallest additional co-activation pressure while respecting the per-instance capacity constraints. 
We provide the formal optimization and full algorithm in Appendix~\ref{appendix:placement}.

\section{Implementation}
\label{sec:implementation}

We implement \SysName on top of SGLang~\cite{sglang_git} with about 4K lines of Python and 300 lines of CUDA/C++ code, extending SGLang to support disaggregated MoE inference. 
On the attention side, \SysName reuses SGLang's request batching, dispatching, and KV-cache management. 
For cross-sub-cluster communication, \SysName implements the adaptive two-phase mechanism with NVSHMEM~\cite{nvshmem} and GPUDirect RDMA, while intra-node collectives over NVLink are implemented using NCCL. 
Specifically, \SysName uses NVSHMEM's one-sided \texttt{putmem\_signal}/\texttt{signal\_wait} primitives to directly write payloads into receiver GPU memory and signal completion. 
We pack lightweight metadata, including layer index and token count, into the same signal value to avoid separate metadata transfers; CPU-side metadata unpacking is performed only at the first MoE layer and then reused for subsequent layers. 
We also tune NVSHMEM parameters, including IBGDA transport, request-batching threshold, and per-peer RC queue count, for our communication pattern.
We place the shared expert on the attention side and execute it while each attention instance transfers intermediate data to the MoE side and waits for the results, thereby overlapping communication with computation. 
On the MoE side, each MoE instance runs AEBS (Algorithm~\ref{algo:aebs}) as a GPU kernel.

\section{Evaluation}
\label{sec:evaluation}

\subsection{Experimental Setup}
\label{sec:setup}

\subtitle{Testbed.}
We deploy \SysName on a GPU cluster of up to 4 nodes. Each node has 128 CPU cores, 2 TB of host memory, and 8 NVIDIA H100 GPUs, each with 80~GB memory. Each GPU is connected with a 400 Gbps InfiniBand NIC. GPUs within a node are interconnected via 900 GB/s NVLink.

\begin{figure*}[t]
    \centering
    \includegraphics[width=\textwidth]{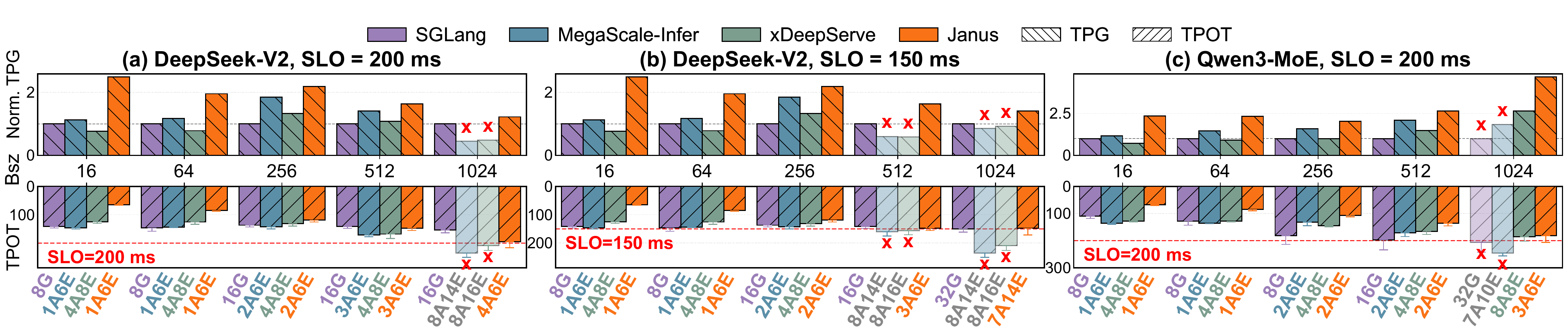}
    \caption{TPOT and normalized per-GPU throughput across batch sizes. 
    (a) DeepSeek-V2, SLO = 200\,ms. 
    (b) DeepSeek-V2, SLO = 150\,ms. 
    (c) Qwen3-MoE, SLO = 200\,ms. 
    Annotations (e.g., \texttt{1A6E}) denote the configurations selected by disaggregated systems, while \texttt{$X$G} ($X$ total GPUs) denotes the configuration used by \textit{SGLang}. The red dashed line marks the SLO threshold, and the whisker marker shows the P99 TPOT.}
    \label{fig:dsv2_e2e_scale}
    \vspace{-.2in}
\end{figure*}

\subtitle{Models and workloads.}
We evaluate \SysName on representative large MoE models, including DeepSeek-V2~\cite{deepseekv2}, Qwen3-MoE~\cite{qwen3}, and Scaled-DS, a family of scaled DeepSeek-style variants that stress different routing and expert configurations.
We consider two Scaled-DS variants.
Scaled-DS-1 uses top-$k=8$ routing over 160 experts per layer, with each expert having intermediate size 1024.
Scaled-DS-2 also uses top-$k=8$, but expands the expert pool to 200 experts and increases the per-expert size to 1536.
All model parameters and KV caches are stored in BF16 format.
We use two representative workloads.
First, we replay requests derived from the ShareGPT dataset~\cite{sharegpt2023}, with an average input length of 16 tokens and an average output length of 256 tokens.
Second, we use BurstGPT~\cite{burstGPT_arxiv24} to synthesize realistic dynamic arrivals that mimic production LLM services.

\subtitle{Baselines.}
We compare \SysName against three baselines.

\emph{(1) SGLang:}
We use vanilla SGLang as our monolithic baseline. 
It deploys the entire MoE model as a single instance under a fixed parallelism configuration, forcing attention and MoE components to share the same parallelism degree.
Therefore, SGLang scales only at a coarse granularity, such as deploying the full model on 8, 16, 32, 64 GPUs.

\emph{(2) MegaScale-Infer:}
MegaScale-Infer is a state-of-the-art disaggregated MoE inference system~\cite{megascale}.
We implement it on top of \SysName's codebase as it is not publicly available. 
Specifically, we replace \SysName's AEBS with random expert scheduling, a common strategy used in existing systems including EPLB~\cite{EPLB2024}. 
Unlike \SysName's two-phase communication design, MegaScale-Infer performs gating on the attention side, requiring each attention instance to send activations and metadata to all MoE instances that host activated experts.
Compared with \SysName, MegaScale-Infer also adopts coarser-grained resource scaling: it restricts the resource configuration space to plans that balance attention-side and MoE-side times for pipelined execution.

\emph{(3) xDeepServe:} 
We also implement xDeepServe~\cite{xdeepserve}, an NPU-superpod-based disaggregated inference system, on top of \SysName, as another baseline.
xDeepServe uses EPLB-like expert scheduling. 
For communication, it targets superpod-scale deployments and incurs more extensive cross-node traffic than \SysName, including all-to-all transfers between attention and MoE nodes.
It also performs gating on the attention side.
Since xDeepServe does not provide a resource-scaling policy, we simply scales it in units of 4 GPUs.

\subtitle{Metrics.}
To evaluate \SysName in decode-centric serving scenarios, we use two primary metrics: time per output token (TPOT) and throughput per GPU (TPG).
TPOT captures token-level latency during decoding and is the SLO metric used throughout our evaluation.
TPG measures resource efficiency, computed as the total output-token throughput divided by the number of GPUs used.

\begin{figure}[t]
    \centering
    \includegraphics[width=0.95\linewidth]{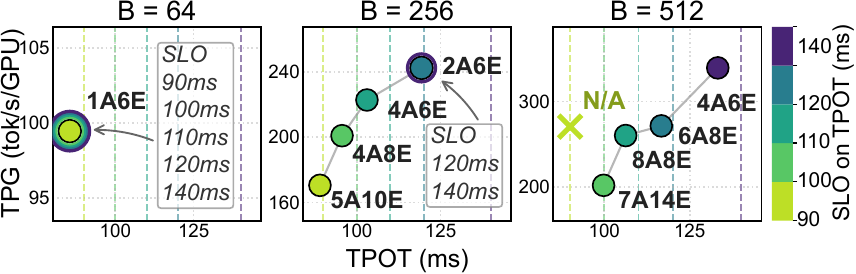}
    \caption{Performance of \SysName under various SLOs.}
    \label{fig:tradeoff}
    \vspace{-.2in}
\end{figure}

\subsection{End-to-End Performance}
\label{sec:e2e}

\subtitle{Per-GPU throughput and SLO attainment.}
We compare the end-to-end performance of \SysName against \textit{SGLang}, \textit{MegaScale-Infer}, and \textit{xDeepServe}.
Fig.~\ref{fig:dsv2_e2e_scale} reports TPOT and normalized per-GPU throughput across batch sizes using DeepSeek-V2 and Qwen3-MoE under 200\,ms and 150\,ms TPOT SLOs.
\SysName consistently satisfies the target SLOs across all evaluated batch sizes and models.
In contrast, \textit{MegaScale-Infer} and \textit{xDeepServe} violate the SLO on DeepSeek-V2 as the batch size increases, with violations appearing at batch size 512 under the 150\,ms SLO and at batch size 1024 under the 200\,ms SLO.
\textit{SGLang} also fails to satisfy the SLO on Qwen3-MoE at batch size 1024.

Fig.~\ref{fig:dsv2_e2e_scale} further shows that \SysName achieves substantially higher resource efficiency through fine-grained, module-specific scaling.
By independently selecting the numbers of attention and expert instances, \SysName improves per-GPU throughput by up to 4.7$\times$, 2.2$\times$, and 3.3$\times$ over \textit{SGLang}, \textit{MegaScale-Infer}, and \textit{xDeepServe}, respectively.
These gains come from avoiding the coarse provisioning decisions made by existing systems.
Under light load, \textit{SGLang} and \textit{xDeepServe} over-provision attention capacity because they scale only in coarse units, whereas \SysName uses compact asymmetric configurations such as \texttt{1A6E} and allocates most resources to the MoE side.
As load increases, \SysName incrementally adds attention capacity to avoid attention-side bottlenecks; under tighter SLOs, it also expands the MoE side to reduce the maximum activated-expert count $a_{\max}$.
In contrast, \textit{MegaScale-Infer} and \textit{xDeepServe} are limited by coarser configuration spaces and less effective expert scheduling.
Overall, \SysName improves both SLO attainment and per-GPU throughput by matching the resource allocation to the distinct scaling behavior of attention and MoE layers.

\subtitle{Performance under various SLOs.}
Fig.~\ref{fig:tradeoff} stress-tests \SysName under different TPOT SLOs and batch sizes.
The selected configuration changes substantially with the latency target, demonstrating the need for SLO-aware resource scaling.
At a small batch size of $B=64$, \texttt{1A6E} already satisfies all evaluated SLOs and achieves about 99 tok/s/GPU, indicating that additional resources would provide little benefit in this regime.
At $B=256$, relaxing the SLO allows \SysName to move from more heavily provisioned configurations such as \texttt{5A10E} to more resource-efficient ones such as \texttt{2A6E}, increasing TPG from roughly 170 to 240 tok/s/GPU.
At $B=512$, the strictest SLO is infeasible, while looser SLOs enable progressively higher-throughput configurations; in particular, \texttt{4A6E} achieves about 340 tok/s/GPU under the most relaxed SLO.
These results expose a clear latency--throughput trade-off: tighter SLOs require more conservative resource allocations to reduce TPOT, whereas relaxed SLOs allow \SysName to use fewer GPUs and maximize per-GPU throughput.

\begin{figure}[t]
    \centering
    \includegraphics[width=0.95\linewidth]{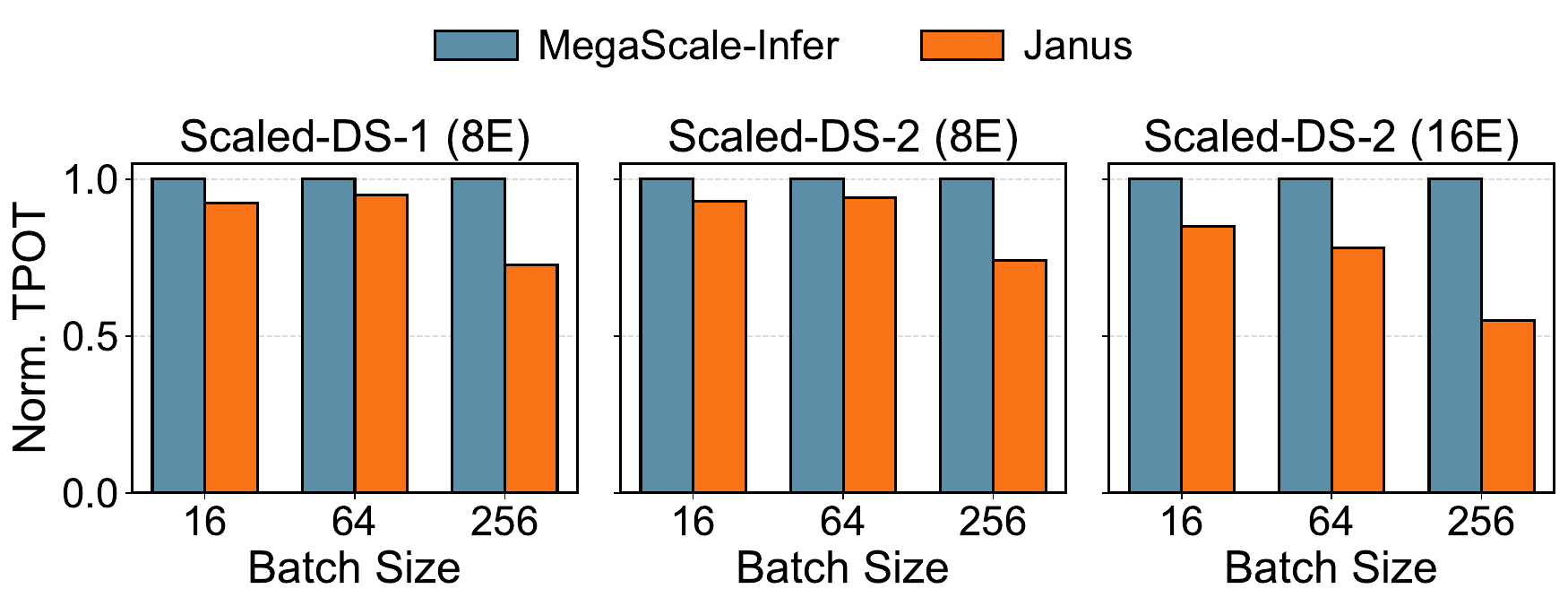} 
    \caption{Normalized TPOT under various model variants.}
    \label{fig:scaleMoE}
    \vspace{-.15in}
\end{figure}

\subtitle{Performance on model variants.} 
Fig.~\ref{fig:scaleMoE} compares \SysName with MegaScale-Infer on Scaled-DS variants.
For Scaled-DS-1 with 8 MoE instances, \SysName reduces TPOT more at larger batch sizes, where its adaptive two-phase communication better amortizes cross-node transfer overhead.
For Scaled-DS-2, 8 MoE instances leave little replica redundancy and limit scheduling gains.
Scaling to 16 MoE instances restores redundancy, enabling \SysName to combine communication efficiency with AEBS and reduce TPOT by 41--50\%.

\begin{figure}[t]
    \centering
    \includegraphics[width=0.9\linewidth]{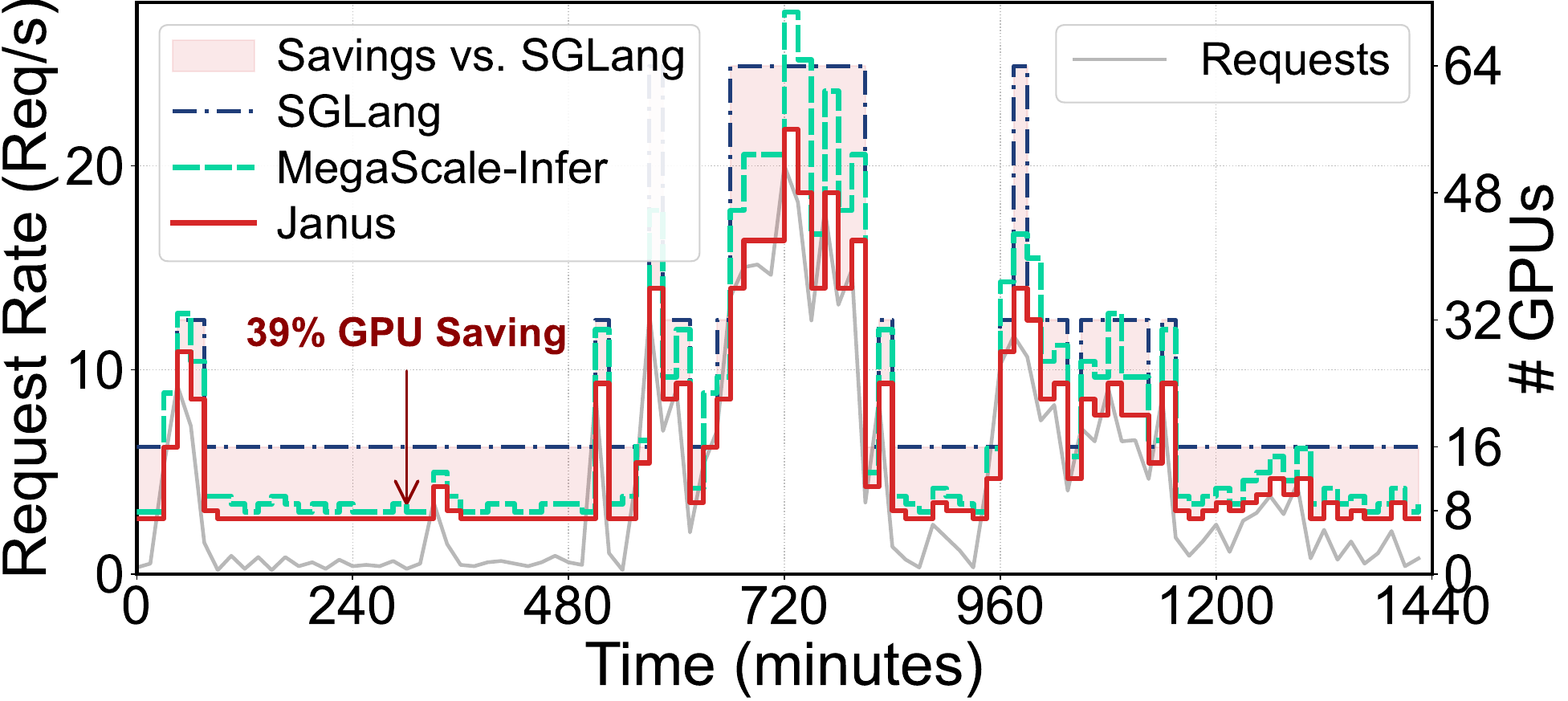}
    \caption{Scaling behaviors over a 24-hour production trace with a 15-minute decision interval. \SysName tracks load changes with fine-grained scaling, while SGLang over-provisions by snapping to 16/32/64-GPU tiers and MegaScale-Infer uses more GPUs due to its coarser scaling policy.}
    \label{fig:trace_scaling}
    \vspace{-.15in}
\end{figure}


\subtitle{Scaling under real-world workloads.} 
Fig.~\ref{fig:trace_scaling} evaluates \SysName under a 24-hour production trace with a 15-minute scaling interval, and compares it with SGLang and MegaScale-Infer.
Since continuously running all systems over the full trace would require substantial cluster time, we evaluate scaling behavior through trace-driven simulation using the measured performance of various systems.
\SysName closely tracks the diurnal load by continuously adjusting the numbers of attention and MoE instances, scaling between 7 and 56 GPUs.
In contrast, SGLang can only snap to coarse 16/32/64-GPU tiers, leading to over-provisioning during low-load periods.
MegaScale-Infer also allocates more GPUs than \SysName because its coarser-grained scaling policy can skip configurations that are more resource-efficient.
As a result, \SysName reduces GPU-hour consumption by 39\% compared with SGLang and by 16\% compared with MegaScale-Infer, while maintaining the target latency requirements.
\subsection{Microbenchmarks}
\label{sec:ablation}



\begin{figure}[t]
    \centering
    \includegraphics[width=0.95\linewidth]{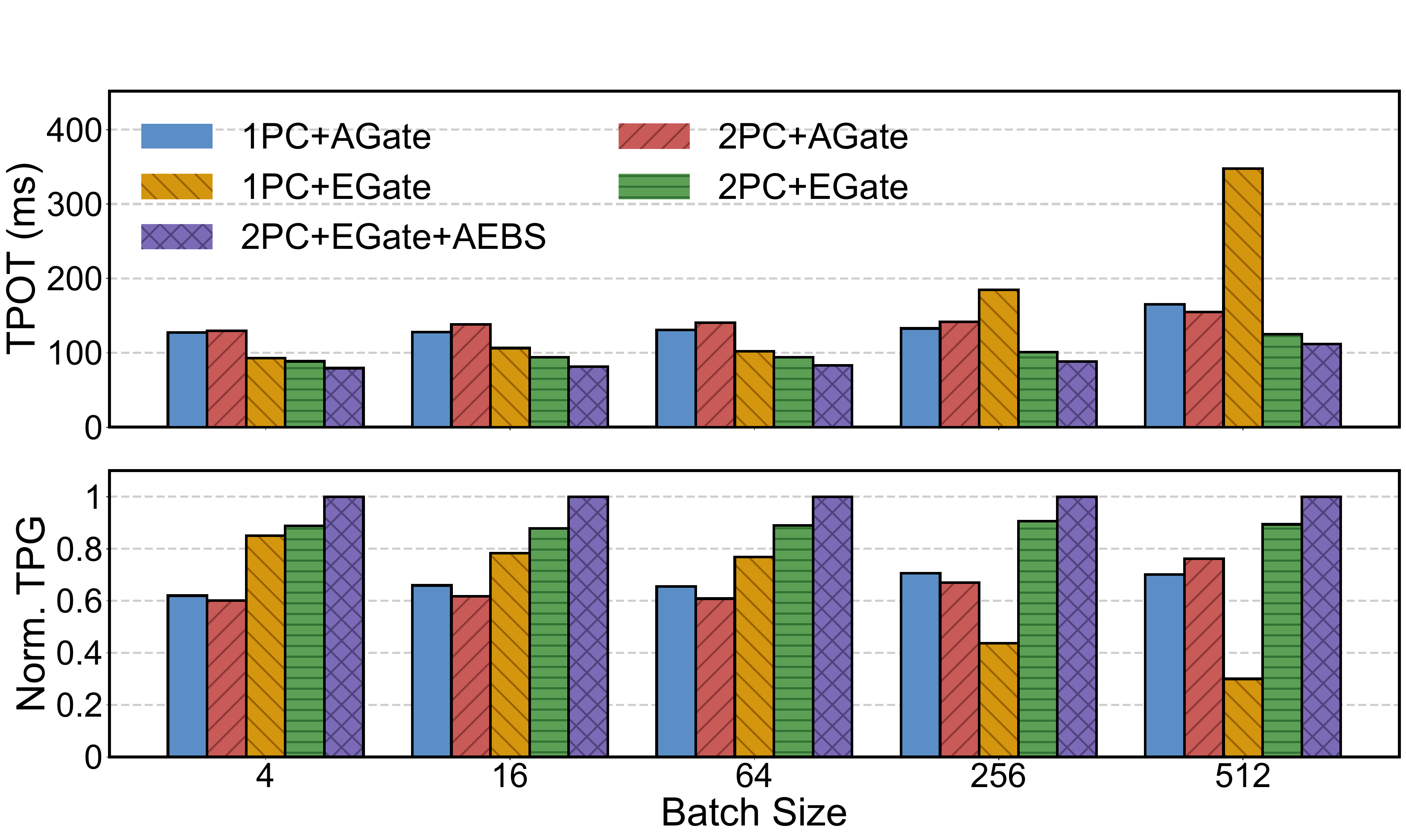}
    \caption{Performance breakdown of \SysName's designs.}
    \label{fig:ablation}
    \vspace{-.2in}
\end{figure}



\subtitle{Performance breakdown.}
We ablate three mechanisms in \SysName: cross-sub-cluster communication, gating location, and activated-expert-balanced scheduling (AEBS).
Here, \textit{1PC} and \textit{2PC} denote one-phase and two-phase communication, while \textit{AGate} and \textit{EGate} denote attention-side and MoE-side gating, respectively.
The full \SysName uses \textit{2PC+EGate+AEBS}.
Fig.~\ref{fig:ablation} reports TPOT and normalized throughput across batch sizes.
The results show that MoE-side gating must be paired with two-phase communication.
Without intra-node aggregation, \textit{1PC+EGate} sends ungated activations directly across nodes, increasing TPOT to 185\,ms and 350\,ms at batch sizes 256 and 512, with throughput dropping to 44\% and 30\% of the full \SysName. 
With \textit{2PC}, intra-node aggregation reduces cross-node transfers and makes \textit{EGate} consistently effective.
As a result, \textit{2PC+EGate} improves throughput over \textit{2PC+AGate} by 4--34\%, since it also avoids sending top-$k$ routing metadata on each attention-to-MoE link.
Finally, adding \textit{AEBS} further improves throughput by 11--15\% by reducing MoE stragglers.

\begin{figure}[t]
    \centering
    \includegraphics[width=0.95\linewidth]{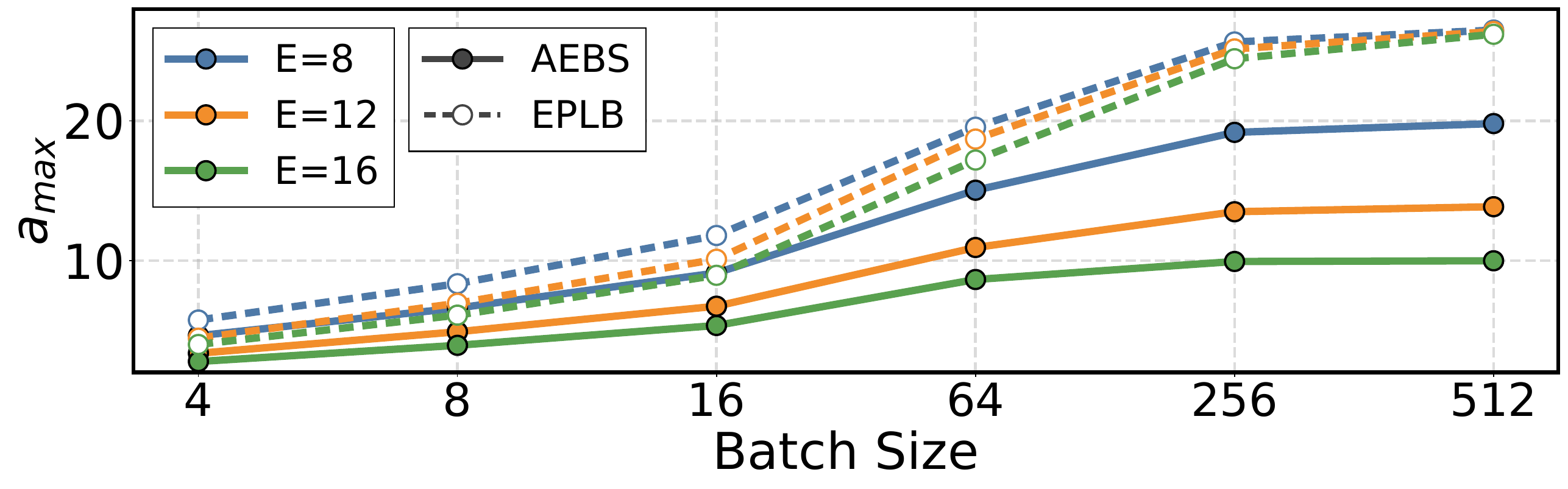}
    \caption{Maximum activated-expert count $a_{\max}$ under different batch sizes and MoE-side scales (E).}
    \label{fig:activation_max}
    \vspace{-.2in}
\end{figure}

\begin{figure}[t]
    \centering
    \includegraphics[width=0.95\linewidth]{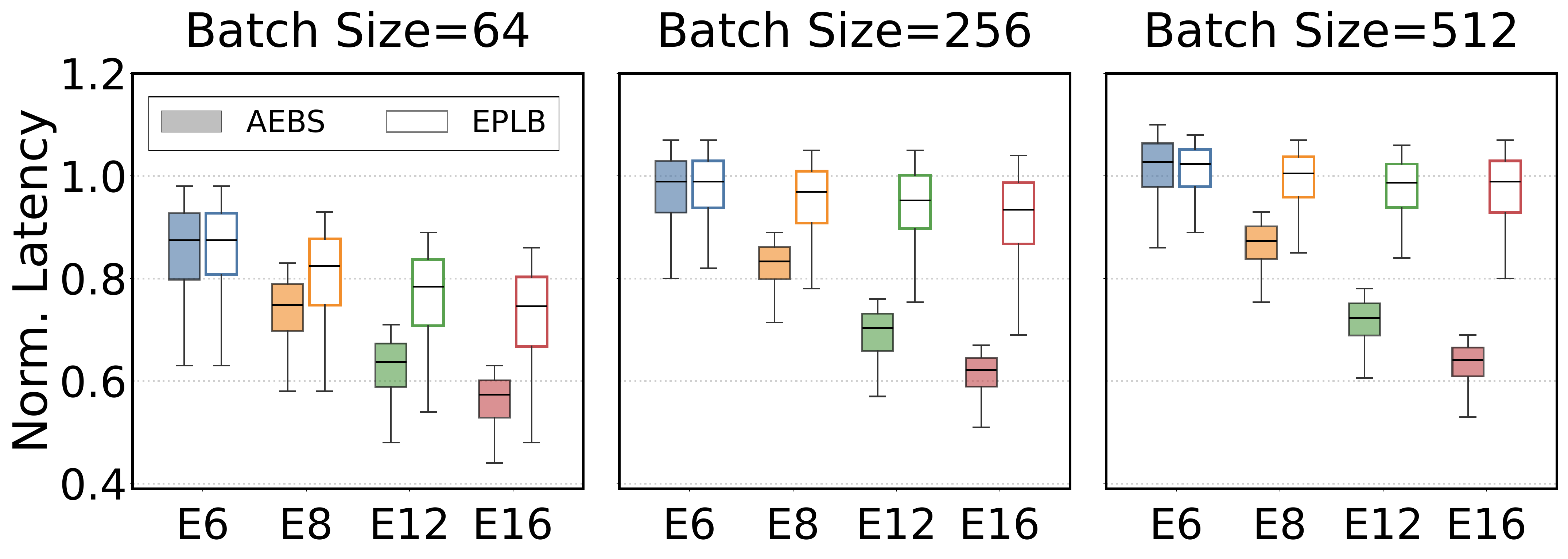}
    \caption{MoE-layer inference latency for three cases.}
    \label{fig:ep_latency_box}
    \vspace{-.2in}
\end{figure}

\subtitle{Effects of \SysName's AEBS.}
Figs.~\ref{fig:activation_max} and~\ref{fig:ep_latency_box} evaluate the effectiveness of \SysName's activated-expert-balanced scheduling (AEBS).
Fig.~\ref{fig:activation_max} reports the maximum number of activated experts assigned to any MoE instance under different batch sizes and MoE-side scales.
This metric captures the straggler effect in MoE execution, since each layer must wait for the instance with the largest activated-expert count.
Compared with EPLB, a common expert-parallel scheduling scheme, AEBS consistently reduces the maximum activated-expert count across all batch sizes.
The reduction becomes larger as the number of MoE instances increases from 8 to 16, because higher expert redundancy gives AEBS more choices when distributing activations across instances.

Fig.~\ref{fig:ep_latency_box} shows the resulting MoE-layer latency.
Across batch sizes 64, 256 and 512, \SysName outperforms EPLB in most configurations, with larger gains when more MoE instances are available. For example, at larger batch sizes, increasing the MoE-side scale from E8 to E16 allows \SysName to substantially reduce latency, while EPLB remains close to the baseline latency because it does not explicitly minimize the maximum activated-expert count.
These results confirm that \SysName's scheduling reduces MoE stragglers by balancing activated experts, effectively leading to latency improvements.

\begin{figure}[t]
    \centering
    \includegraphics[width=0.95\linewidth]{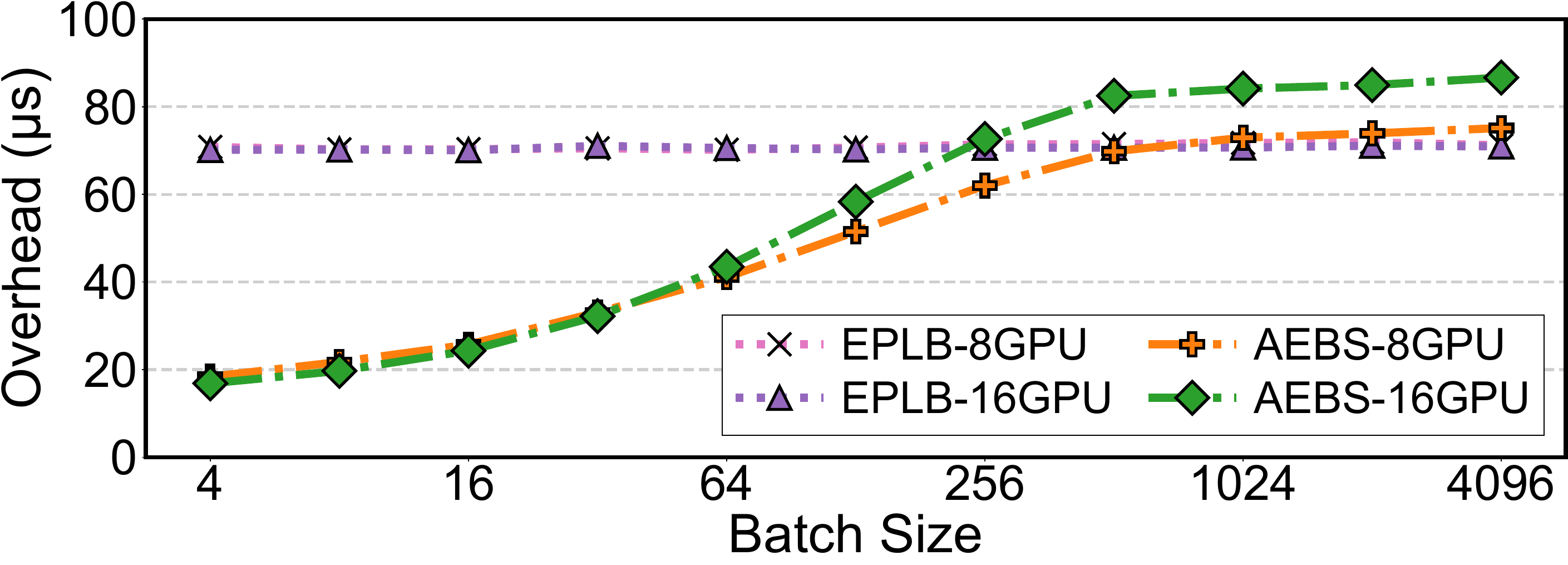}
    \caption{Overhead of \SysName's AEBS.}
    \label{fig:sched_overhead}
    \vspace{-.2in}
\end{figure}

\subtitle{Overhead of \SysName's AEBS.}
Fig.~\ref{fig:sched_overhead} reports the scheduling overhead of AEBS and EPLB under different batch sizes and MoE sub-cluster scales.
AEBS incurs low overhead across all settings: it starts below 20\,\textmu s at small batch sizes and remains below 90\,\textmu s even at batch size 4096.
Its cost increases with batch size because larger batches activate more distinct experts, but gradually plateaus once most experts have been activated.
Increasing the MoE sub-cluster from 8 to 16 GPUs adds only a small overhead, showing that AEBS scales well with the number of MoE instances.
Overall, AEBS introduces negligible scheduling cost and meets the latency requirements of layer-wise MoE execution.

\begin{figure}[t]
    \centering
    \includegraphics[width=0.95\linewidth]{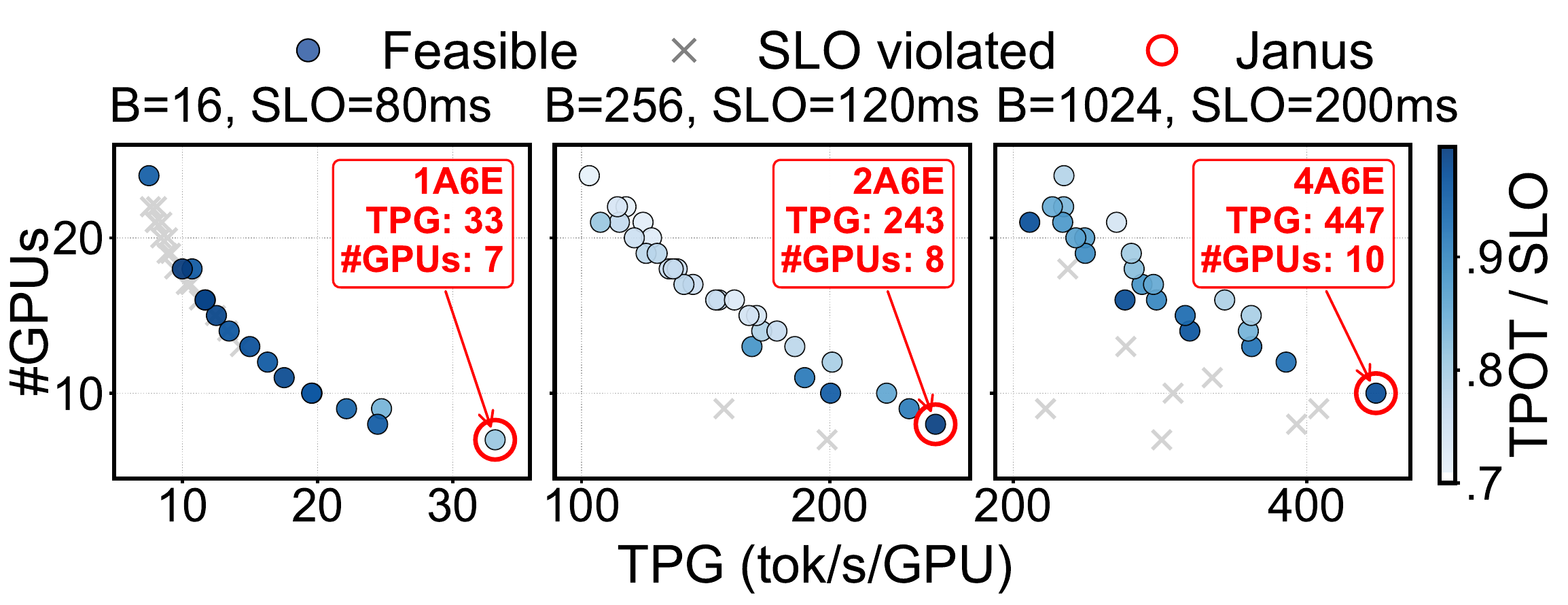}
    \caption{Scaling-policy search space under three representative cases. Each marker denotes a resource configuration $(n_a,n_e)$, plotted by TPG (higher is better) and total GPU count $n_a+n_e$ (lower is better). Circles are SLO-feasible configurations, crosses violate the SLO, and the red ring marks \SysName's selected configuration.}
    \label{fig:solver_micro}
    \vspace{-.2in}
\end{figure}

\subtitle{Effects of \SysName's scaling quality.}
Fig.~\ref{fig:solver_micro} visualizes the $(n_a,n_e)$ search space explored by \SysName across three representative batch-size/SLO settings.
Each marker denotes a candidate resource configuration, plotted by TPG and total GPU count $n_a+n_e$; SLO-feasible configurations are shown as circles and colored by their TPOT/SLO ratio.
The results show that efficient scaling requires searching asymmetric attention/MoE allocations rather than simply adding  GPUs or scaling both sides proportionally.
In all cases, \SysName selects \texttt{1A6E}, \texttt{2A6E}, and \texttt{4A6E}, which satisfy the SLO while achieving high TPG with only 7--10 GPUs.
This confirms that the scaling policy can identify resource-efficient configurations.

\section{Discussion and Other Related Work}
\label{sec:discussion}

\subtitle{Heterogeneous hardware.}
Modern data centers increasingly mix different GPU generations and specialized AI accelerators~\cite{mo2025hetis,helix}. 
Recent inference hardware follows the same trend: for example, NVIDIA's Vera Rubin platform pairs Rubin GPUs for compute-intensive prefill with LPX accelerators optimized for bandwidth-intensive FFN/MoE decode execution. 
\SysName can naturally support such environments by mapping attention and MoE instances to separate hardware pools, and its core mechanisms remain applicable.

\subtitle{Pipelining across attention and MoE.} 
Pipelining attention and MoE execution can improve resource utilization by overlapping the two modules across micro-batches~\cite{megascale}.
However, its benefit is limited without careful design.
Our measurements show that, for typical online batch sizes (often below 100), splitting a batch into multiple micro-batches provides little per-micro-batch latency benefit, while introducing extra synchronization overhead, implementation complexity, and resource interference, consistent with observations from other work~\cite{pan2025efficient,liu2026revealing}.
Effective pipelining therefore requires careful coordination of micro-batch sizing and task scheduling, which is complementary to \SysName.

\section{Conclusion}
\label{sec:conclusion}



We presented \SysName, a scalable MoE inference system that disaggregates attention and MoE layers into separate GPU sub-clusters and scales them independently.
Through adaptive two-phase communication, activated-expert-balanced scheduling, and fine-grained resource scaling, \SysName delivers low-latency, resource-efficient MoE inference under dynamic workloads.
Evaluation shows that \SysName satisfies TPOT SLOs and improves per-GPU throughput by up to 4.7$\times$ over state-of-the-art systems.

\bibliographystyle{plain}
\bibliography{references}

\clearpage   
\appendix

\section{Theoretical Bound for $a_{\max}$}
\label{appendix:maxa}

This appendix derives the closed-form upper bound on $a_{\max}(n_e, B)$ used in \S\ref{sec:exper_mana}. We model expert activation as a \emph{balls-into-bins} process and take an \emph{adversarial view} of AEBS: every activation of a replicated expert is assumed to land on the instance being analyzed, so the bound is independent of the scheduler's routing decisions.

Consider a batch of $B$ tokens, each selecting $K$ experts with activation probabilities $\sum_e p_e = K$, and let $X_e \in \{0,1\}$ denote the event that expert $e$ is hit by at least one token in the batch; then $\Pr(X_e = 1) = 1 - (1 - p_e)^B$. Under the adversarial view, the load on instance $g$ satisfies $a_g \le \sum_{e \in P(g)} X_e$, and expectation gives:
\begin{equation}
\label{eq:expected_act}
\mathbb{E}[a_g] \;\le\; \sum_{e \in P(g)} \bigl[1 - (1 - p_e)^B \bigr].
\end{equation}
Under uniform activation ($p_e = K/E$), this simplifies to $C \cdot [1 - (1 - K/E)^B]$, which grows with $B$ and saturates at $C$ as every hosted expert becomes almost surely activated. The bottleneck instance is the one that attains $\bar a_{\max} := \max_g \mathbb{E}[a_g]$.

Although top-$K$ gating couples the indicators $\{X_e\}$, the resulting coupling is \emph{negatively} associated, which preserves $\mathrm{Var}(a_g) \le \mathbb{E}[a_g]$; applying a Bernstein-type tail bound on each instance and a union bound over $n_e$ instances yields:
\begin{equation}
\label{eq:max_a_bound}
a_{\max}(n_e, B) \;\le\; \left\lfloor \min\!\Bigl(\, C,\;\; \bar a_{\max} + \sqrt{2\,\bar a_{\max}\,\ln n_e}\,\Bigr) + 1 \right\rfloor.
\end{equation}
$a_{\max}$ counts distinct experts and is integer-valued; the $+1$ slack absorbs the replication-induced overflow that lets the bottleneck instance occasionally host $\lceil E/n_e \rceil + 1$ distinct experts.

Two regimes follow. When $B$ is small, $\bar a_{\max} \ll C$ and $a_{\max}$ grows with $B$, driving $T_{\mathrm{moe}}^{(\ell)}$ upward. When $B$ is large, $\bar a_{\max} \to C$ and $a_{\max}$ plateaus, so $T_{\mathrm{moe}}^{(\ell)}$ is effectively capped while $T_{\mathrm{attn}}^{(\ell)}$ continues to grow with $B/n_a$; this explains the diminishing returns of batch size on end-to-end throughput. Eq.~\eqref{eq:max_a_bound} is conservative because it treats activations adversarially and ignores the variance-flattening effect of replication and placement (\S\ref{sec:exper_mana}) as well as the peak-reduction effect of AEBS (\S\ref{sec:lb_scheduling}); the Monte Carlo estimator $\widehat{a}_{\max}$ used by the scaling solver absorbs these effects.

\begin{figure}[t]
    \centering
    \includegraphics[width=.9\linewidth]{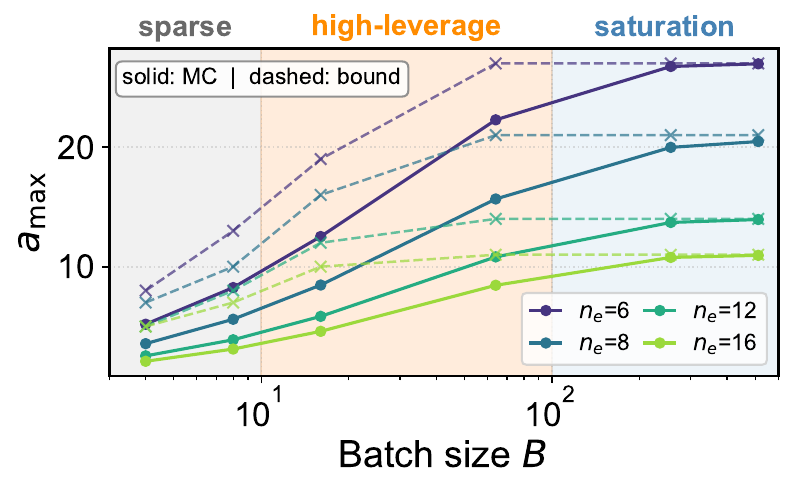}
    \caption{Analytical bound (dashed) vs.\ Monte Carlo estimate (solid) on ShareGPT across $n_e \in \{6,8,12,16\}$, with three batch-size regimes shaded. The high-leverage window $B\!\in\![10,100]$ is where $a_{\max}$ is simultaneously most sensitive to placement (steepest slope) and already at $30\!-\!60\%$ of $C$, and it coincides with the per-instance batch sizes reported in online decode traces.}
    \label{fig:regime_overlay}
\end{figure}

\PHM{Empirical validation and high-leverage regime.}
Figure~\ref{fig:regime_overlay} overlays the analytical bound against the layer-averaged Monte Carlo estimate $\widehat{a}_{\max}$ on ShareGPT across $n_e \in \{6, 8, 12, 16\}$, with three batch-size regimes shaded. \emph{The bound holds on all cells}: in the saturation regime ($B \ge 64$) the gap is within one or two experts, and even at small $B$ it stays below $\sim$2$\times$. Two observations justify treating this as an acceptable, usefully conservative bound rather than a loose one. First, the gap is one-sided---the bound never under-predicts, so using it in the scaling solver can only err on the side of over-provisioning, which is the safe direction under SLO constraints. Second, the predicted values remain within the range of activated-expert counts actually observed in practice: the MoE latency measurements in Fig.~\ref{fig:activation_pattern} span roughly 8--22 activated experts, and all bound values in Fig.~\ref{fig:regime_overlay} fall inside or below this envelope, so the bound does not drive the solver into regions that contradict measured behavior. The residual looseness reflects the adversarial assumption that ignores the variance-flattening effect of replication and placement and the peak-reduction effect of AEBS; $\widehat{a}_{\max}$ closes this gap at decision time by incorporating both effects.

Beyond validating the bound, Fig.~\ref{fig:regime_overlay} reveals three regimes with sharply different scheduling leverage. (i) \emph{Sparse} ($B \lesssim 10$, gray band): $\widehat{a}_{\max}$ is small ($\le 4$ across all $n_e$) and largely insensitive to placement---there are too few tokens for policy to matter. (ii) \emph{Saturation} ($B \gtrsim 100$, blue band): $\widehat{a}_{\max}$ plateaus near $\min(C, E/n_e)$ (e.g., $n_e{=}6$: $19.98\!\to\!20.47$ from $B{=}256$ to $B{=}512$; $n_e{=}16$: $10.78\!\to\!10.95$); the ceiling is structural and no scheduling policy can push $\widehat{a}_{\max}$ below it. (iii) \emph{High-leverage} ($B \in [10,100]$, orange band): the curve exhibits its steepest slope---each $4\times$ increase in $B$ raises $\widehat{a}_{\max}$ by $4\!-\!7$ experts (e.g., $n_e{=}6$: $B{=}16\!\to\!64$ moves $\widehat{a}_{\max}$ from $8.46$ to $15.66$; $n_e{=}12$: $5.84\!\to\!10.82$), and absolute values already reach $30\!-\!60\%$ of $C{=}27$. Because $T_{\mathrm{moe}}^{(\ell)} = \beta^{(\ell)} \cdot a_{\max}^{(\ell)} + c_e^{(\ell)}$ dominates per-layer latency once $a_{\max}$ is in this range, a 2--3-expert shift in placement suffices to move TPOT across the SLO. Per-instance batch sizes in online decode traces~\cite{splitwise,llmtrace2024} sit inside exactly this window, which motivates concentrating replication, placement, and AEBS on $B \in [10,100]$ rather than on sparse or saturated regimes.

\section{Activation-Aware Replica Placement}
\label{appendix:placement}

This appendix provides the details of \SysName's expert placement in \S\ref{sec:exper_mana}.

\paragraph{Replica count.}
Given $n_e$ MoE instances each with $C$ expert slots, the $S = n_e \cdot C$ total slots first seat one replica of each of the $E$ logical experts; the remaining $S - E$ slots provide redundancy. \SysName assigns these redundant slots iteratively: using activation counts $c(e)$ over a sliding window, it repeatedly picks the expert with the largest per-replica load $l(e) = c(e) / R(e)$ and grants it one more replica, until all $S - E$ extra slots are exhausted. Hot experts accumulate more replicas, cold experts remain singleton, and per-replica activation pressure is equalized.

\paragraph{Placement optimization.}
Given replica assignments $\{R(e)\}_{e=1}^{E}$ and per-instance capacity $C$, let $x_{e,g} \in \{0,1\}$ indicate whether a replica of logical expert $e$ is placed on instance $g$, and let $a(e,e')$ denote the co-activation frequency between logical experts $e$ and $e'$ estimated from recent traces. We define the co-activation load on instance $g$ as:

\begin{equation}
\label{eq:co_activated}
I(g) \;=\; \sum_{\substack{e,e' \in P(g)\\ e<e'}} a(e,e')
\end{equation}

Colocating experts with high $a(e,e')$ raises concurrent activations on that instance and thus MoE latency. \SysName solves the min--max assignment:

\begin{equation}
\label{eq:obj_placement}
\begin{aligned}
\min_{\{x_{e,g}\}} \quad & \max_{g \in \{1,\dots,n_e\}} I(g) \\[0.4em]
\text{s.t.} \quad
& \sum_{e=1}^{E} x_{e,g} \le C, \\
& \sum_{g=1}^{n_e} x_{e,g} = R(e), \\
& x_{e,g} \in \{0,1\}.
\end{aligned}
\end{equation}

Eq.~\eqref{eq:obj_placement} reduces to unrelated-machines scheduling and is NP-hard~\cite{lenstra1990unrelated}. \SysName uses the greedy heuristic in Algorithm~\ref{algo:placement}. It first initializes per-instance placement sets, remaining slots, and a bitmap recording whether an instance already hosts a replica of a given logical expert (lines~1--3). It then iterates over replicas in descending order of load: if there exists an instance with free capacity that does not yet host that expert, the replica is placed on the instance that adds the least co-activation penalty (lines~5--10). Otherwise, a bounded swap between two instances is performed to create a feasible placement with minimal incremental co-activation cost (lines~11--18). This heuristic closely approximates the min--max objective while remaining efficient enough for periodic online reconfiguration.

\begin{algorithm}[t]
\caption{Activation-Aware Replica Placement}
\small
\label{algo:placement}
\textbf{Input:} \\
-- $n_e$: number of instances, $C$: capacity per instance \\
-- $\mathcal{R}$: set of replicas, $l_i$: load of replica $i$, $e_i$: logical expert of replica $i$ \\
\textbf{Output:} \\
-- $P(g)$: replicas assigned on instance $g$

\begin{algorithmic}[1]

\State Initialize $P(g)\gets\emptyset$, $slots[g]\gets C$ for all $g\in \{1,2,\dots,n_e\}$
\State Initialize $x_{e,g}\gets 0$ for all experts $e$ and $g$

\State Sort replicas $\mathcal{R}$ in decreasing order of $l_i$

\ForAll{$i \in \mathcal{R}$}

    \State $G_i \gets \{\, g \in G \mid slots[g] > 0 \land x_{e_i,g}=0 \,\}$

    \If{$G_i \neq \emptyset$} \Comment{Slots feasible}
        \State $g^* \gets \arg\min_{g \in G_i} \sum_{j \in P(g)} a(i,j)$
        \State $P(g^*) \gets P(g^*) \cup \{e_i\}$
        \State $slots[g^*] \gets slots[g^*] - 1$
        \State $x_{e_i,g^*} \gets 1$

    \Else \Comment{No feasible slot; resolve via swapping}
        \State $G_i^{\neg} \gets \{\, g \in G \mid x_{e_i,g} = 0 \,\}$ \Comment{Instances without expert $e_i$}
        \State $H_i \gets \{\, h \in G \mid slots[h] > 0 \,\}$ \Comment{Instances with free slots}
        \State Find $g \in G_i^{\neg}$, $h \in H_i$, and $j \in P(g)$ such that $x_{e_j,h} = 0$
        \Statex \textit{\# Minimize co-activation load penalty of swapping}
        \State $(g,j,h) \gets \arg\min_{g,j,h} \Delta I(i \to g,\; j \to h)$

        \Statex \textit{\#  Apply replica swapping}
        \State $P(g) \gets (P(g)\setminus\{j\}) \cup \{i\}$
        \State $P(h) \gets P(h) \cup \{j\}$
        \State $x_{e_j,g} \gets 0,\quad x_{e_j,h} \gets 1,\quad x_{e_i,g} \gets 1$

    \EndIf
\EndFor

\end{algorithmic}
\end{algorithm}

\end{document}